\documentclass[letterpaper,aps,twocolumn,showpacs,superscriptaddress,floats,prb]{revtex4}
\usepackage{amsmath}
\usepackage{graphicx,epsfig,dsfont}
\usepackage{amssymb}
\usepackage[dvips]{color}
\usepackage{ulem}

\newcommand{\be}{\begin{equation}}
\newcommand{\ee}{\end{equation}}
\newcommand{\bea}{\begin{eqnarray}}
\newcommand{\eea}{\end{eqnarray}}

\newcommand{\nn}{\nonumber}
\newcommand{\pdag}{{\phantom{\dagger}}}

\begin{document}

\title{Dislocation-mediated melting of one-dimensional Rydberg crystals}

\author{Eran Sela}
\affiliation{Institut f\"ur Theoretische Physik, Universit\"at zu K\"oln,
Z\"ulpicher Str. 77, 50937 K\"oln, Germany
}
\author{Matthias Punk}
\affiliation{Physik Department, Technische Universit{\"a}t M{\"u}nchen, James-Franck-Strasse, 85748 Garching, Germany}
\affiliation{Department of Physics, Harvard University, Cambridge MA 02138, USA}

\author{Markus Garst}
\affiliation{Institut f\"ur Theoretische Physik, Universit\"at zu K\"oln,
Z\"ulpicher Str. 77, 50937 K\"oln, Germany
}

\begin{abstract}
We consider cold Rydberg atoms in a one-dimensional optical lattice in the Mott regime with a single atom per site at zero temperature. An external laser drive with Rabi frequency $\Omega$ and laser detuning $\Delta$,
creates Rydberg excitations whose dynamics is governed by an effective spin-chain model with (quasi) long-range interactions. This system possesses intrinsically a large degree of frustration resulting in a ground-state phase diagram in the
$(\Delta,\Omega)$ plane with a rich topology. As a function of $\Delta$, the Rydberg blockade effect gives rise to a series of crystalline phases commensurate with the optical lattice that form a so-called devil's staircase. The Rabi frequency, $\Omega$, on the other hand, creates quantum fluctuations that eventually lead to a quantum melting of the crystalline states. Upon increasing $\Omega$, we find that generically a commensurate-incommensurate transition to a floating Rydberg crystal occurs first, that supports gapless phonon excitations. For even larger $\Omega$, dislocations within the floating Rydberg crystal start to proliferate and
a second, Kosterlitz-Thouless-Nelson-Halperin-Young dislocation-mediated melting transition finally destroys the crystalline arrangement of Rydberg excitations. This latter melting transition is generic for one-dimensional Rydberg crystals and persists even in the absence of an optical lattice. The floating phase and the concomitant transitions can, in principle, be detected by Bragg scattering of light.
\end{abstract}

\date{\today}

\maketitle

 \section{Introduction}

The experimental progress in manipulating ultracold atomic gases has established these systems by now as versatile quantum simulators of basic many-body Hamiltonians with an unprecedented control over coupling parameters.\cite{BDZ} The interactions between ultracold neutral atoms are, however, usually limited to on-site- or contact-interactions due to the s-wave nature of ultracold collisions. In order to overcome this limitation, there has been a considerable effort recently to create atomic gases with (quasi) long-range dipolar interactions.\cite{Lahaye}  Generally, long-range interactions can give rise to a variety of competing ground states and stabilize new quantum phases of matter, that are unfamiliar from systems with finite-range interactions only. In the context of ultracold gases of dipolar atoms various examples have been studied theoretically so far, e.g., supersolids,\cite{Burnell,Pupillo,Pollet} crystalline phases,\cite{Buechler,Astrakharchik,Hauke} liquid crystal and nematic phases.\cite{Quintanilla,Fregoso,Lin,Dalmonte}

Apart from using atoms with a static dipole moment, long-range forces can also be generated by laser driving atoms to dipolar states or, alternatively, to highly excited Rydberg states. Such excited Rydberg atoms then experience either a dipolar or a van der Waals force, respectively, that can become quite strong due to their huge polarizability that scales as $n^{7}$ where the principal quantum number, $n$, is typically on the order of $n \sim 50$ in current experiments. The interesting aspect of this realization is that in either case
the additional long-range interaction acts only between excited atoms. As a consequence, it becomes energetically unfavorable to excite two atoms at short distance -- a phenomenon known as the dipole-blockade effect.\cite{Lukin01,Tong04,Singer04,Urban09,Gaetan09} In the presence of many atoms, this blockade mechanism gives rise to interesting collective behavior\cite{Heidemann07,Heidemann08,Henkel,Cinti} especially in low-dimensional systems.\cite{Honer,Weimer2008,Low,Olmos,Weimer,Pohl,Schachenmayer,Viteau}

In the present work, we consider a regular one-dimensional (1D) lattice of Rydberg atoms and study the ground-state phase diagram as a function of laser detuning, $\Delta$, and Rabi frequency, $\Omega$. Interestingly, this 1D system possesses intrinsically, as we will argue below, a large degree of frustration, and, as a result, the dipole-blockade leads here to various liquid-like and crystalline phases with the concomitant melting transitions. This system was studied in a number of recent publications\cite{Weimer2008,Low, Olmos,Weimer,Pohl,Schachenmayer,Lesanovsky} that elucidated several aspects of the problem. The density of Rydberg excitations and its dependence on the Rabi frequency $\Omega$  were studied experimentally and theoretically
for a trapped gas without an optical lattice,\cite{Weimer2008,Low} the dynamics of excitations for Rabi frequencies larger than the interaction energy was investigated in Ref.~\onlinecite{Olmos}, and crystalline phases were considered in Refs.~\onlinecite{Weimer,Pohl,Schachenmayer,Lesanovsky}. It is the purpose of this work to complement and to complete the picture of
the phase diagram of the system, to identify the universality classes of the phase transitions, and  to put it into context with well-established results in statistical physics about melting of crystalline phases.\cite{ChaikinBook,Bak81} In particular, we point out that as a function of laser detuning, $\Delta$, the Rydberg excitations first condense into a gapless incommensurate crystalline state in agreement with Ref.~\onlinecite{Weimer}, but we identify the transition to be of Berezinskii-Kosterlitz-Thouless\cite{Kosterlitz73} type in analogy to the Kosterlitz-Thouless-Nelson-Halperin-Young scenario of dislocation-mediated melting of two-dimensional crystals on periodic substrates.\cite{Nelson,Young}  As quantum fluctuations are further suppressed, for example, by lowering $\Omega$ a second, commensurate-incommensurate (C-IC) transition occurs and a Rydberg crystal forms with a periodicity that is commensurate with the optical lattice. At the two transitions, the character of Rydberg correlations changes abruptly which, as we argue, can in principle be detected by Bragg scattering of light. In the following, we introduce the considered model in some detail and discuss the methods applied throughout this work.

\subsection{Model}

We consider ultracold bosonic atoms in a 1D optical lattice in the presence of external laser fields that drive a transition from the ground-state of an atom to a highly-excited Rydberg state. Experimentally, this transition occurs via a two-photon process, the intermediate state of which can be adiabatically eliminated resulting in an effective two-level description in terms of the ground- and the excited Rydberg state.\cite{Atom-PhotonBook} Thus, we start from a Bose-Hubbard model of atoms in a 1D optical lattice together with an additional external driving field at effective Rabi frequency $\Omega$ and detuning $\Delta$. We assign to each atom a pseudospin $\sigma = \downarrow, \uparrow$ that indicates whether it is in the ground- or excited Rydberg state, respectively.
The laser excites Rydberg atoms, and excited atoms can decay by spontaneous emission or collisional processes giving rise to a decay rate $1/\tau$. In general, it is therefore not possible to describe the atomic subsystem by a Hamiltonian but a description in terms of a density operator must be used instead.
However, if we limit ourselves to time-scales much smaller than the life-time $\tau$ of the excited Rydberg atoms we can approximate the evolution of the density operator with the help of an effective Hamiltonian, whose ground state phase diagram can be analyzed.

Employing the rotating wave approximation\cite{Atom-PhotonBook} this Hamiltonian takes the form
\begin{align}
\label{hamHE}
\mathcal H_{\rm HE} =& - \sum_{j,\sigma} t_\sigma \big( b^\dagger_{j \sigma} b^{\ }_{j+1 \sigma} + \text{h.c.} \big)
+ \frac{\Omega}{2} \sum_j \big( b_{j \uparrow}^\dagger b_{j \downarrow}^{\ } +\text{h.c.} \big)
\nn\\
&
- \frac{\Delta}{2} \sum_j (n_{j \uparrow} - n_{j \downarrow})
+ \frac{1}{2} \sum_{j, \sigma} U_{\sigma \sigma} n_{j \sigma} (n_{j\sigma}-1)
\nn \\
& + \sum_{j} U_{\uparrow \downarrow} n_{j \uparrow} n_{j\downarrow}
+ \frac{1}{2}\sum_{j\neq\ell} J^{\ }_{|j-\ell|} \, n_{j \uparrow} n_{\ell\uparrow},
\end{align}
where $b_{j \sigma}$ is the bosonic annihilation operator at lattice site $j$ and $n_{j \sigma} = b^\dagger_{j \sigma} b^{\ }_{j \sigma}$ is the density operator for ground- and excited-state atoms at site $j$.
Generally, the hopping of atoms, $t_\sigma$, and their on-site interaction, $U_{\sigma \sigma'}$, depend on their internal energy state $\sigma$.
In addition, the excited atoms experience a repulsive
interaction denoted by $J_{|r|}$ that decays algebraically with dimensionless distance $r$
\begin{align} \label{LongRangeInt}
J_{|r|} = \frac{J_R}{|r|^\alpha},
\end{align}
where the exponent is either $\alpha = 6$ if the interaction is of van der Waals type or $\alpha = 3$ for a dipolar interaction between excited Rydberg atoms.\cite{Lahaye} Although the interaction of Eq.~\eqref{LongRangeInt} decays rather fast for the exponents of interest, we call it nevertheless (quasi) long-range in this work  in the sense that it is finite for all distances $r$.

We assume that the filling is exactly one atom per site, and we consider the Mott regime where the strong on-site $U \gg t_\sigma$
prohibits the atoms to hop and freezes them to the lattice positions.
Projecting the ``high-energy" Hamiltonian \eqref{hamHE} onto the low-energy Hilbert space with a fixed density of one atom per site one obtains an effective spin Hamiltonian
\begin{align}
\label{Model}
\mathcal H =& \sum_{\ell > j} J_{|\ell-j|} \, \big(S_j^z+1/2\big) \big(S_\ell^z+1/2 \big)
\\\nn&
-\frac{J_\perp}{2} \sum_j \big( S^+_j S^-_{j+1} + \text{h.c.} \big)
-\Delta \sum_j S_j^z + \Omega \sum_j S_j^x,
\end{align}
with $S^\alpha_j$ denoting spin-1/2 operators at lattice site $j$. The Rabi frequency $\Omega>0$ corresponds to a transversal magnetic field and the detuning $\Delta$ is an additional magnetic field component in longitudinal direction. Virtual hopping processes in the Mott regime give rise to a transverse spin interaction $J_\perp \sim t^2/U$ between nearest neighbors; see appendix \ref{app:BEC}. Note that $J_\perp > 0$ is ferromagnetic as we consider bosonic atoms.\cite{Kuklov, Duan}
In appendix \ref{app:BEC} we also shortly discuss the opposite limit of a Bose-Einstein condensate of ground-state atoms and show that the effective Hamiltonian in this case is identified with a spin model similar to Eq.~\eqref{Model} as well.

Eq.~\eqref{Model} is the Hamiltonian that we analyze in this work. In the following, we will use the language of spins and Rydberg atoms interchangeably. For example, the magnetization per site plus one-half, $0 < m+1/2< 1$, corresponds to the mean-number of Rydberg excitations per lattice site.
We will be mainly interested in the ground state phase diagram of the Hamiltonian \eqref{Model} in the $(\Delta,\Omega)$-plane. Note that this phase diagram is symmetric with respect to the line $\Delta= \bar{\Delta} \equiv \sum_{r>0} J_{r}$, which corresponds to vanishing longitudinal magnetic field.
In the following we restrict, however, our discussion to the relevant regime $|\Delta| \ll J_R$, i.e., $|\Delta| \ll \bar{\Delta}$.

In usual experimental settings the Rydberg interaction $J_R$ is by far the largest energy scale,
\begin{align} \label{ParameterRegime}
\Omega, |\Delta|, J_\perp \ll J_R.
\end{align}
Previous studies have been limited to the so-called frozen Rydberg gas limit where the hopping of Rydberg excitations $J_\perp$ is omitted arguing that relevant experimental time-scales
like the Rydberg life-time $\tau$ are usually much smaller than $\hbar/J_\perp$.
Although the bare value of $J_\perp$ is indeed small, it becomes strongly enhanced by renormalization effects generated in second order in $\Omega$.~\cite{Weimer} We therefore find it convenient to keep it explicitly in the model (\ref{Model}) as it simplifies the theoretical analysis.

The analysis of Eq.~\eqref{Model} in the parameter regime \eqref{ParameterRegime} is actually a formidable task  for the following reason.
As a starting point of an analysis, we might neglect the fields, $\Omega$ and $\Delta$, the transverse interaction $J_\perp$ and also all longitudinal interactions between neighbors with a distance larger than a single lattice spacing because of the strong decay of $J_{|r|}/J_R = 1/|r|^\alpha$ with increasing distance $r$. In this case, the Hamiltonian reduces to
\begin{align}
\label{HamCr}
\mathcal{H}_{0} = \sum_{j} J_R \big(S_j^z+1/2\big) \big(S_{j+1}^z+1/2 \big).
\end{align}
It turns out that the ground state of Eq.~\eqref{HamCr} is macroscopically degenerate with zero-point entropy $S = \log \frac{1+\sqrt{5}}{2}$, given by the logarithm of the golden mean,\cite{BaxterBook} as all states without two adjacent up-spins, i.e., two adjacent Rydberg excitations, have the same energy. Almost any additional term will immediately quench this entropy, and it is clear that a perturbative approach in the parameters $\Delta, \Omega, J_\perp$ and $J_{|r|}$ with $r > 1$ is, therefore, inadequate to derive the properties of the Hamiltonian \eqref{Model}.

\subsection{Methods}
One can obtain an idea of the complexity of Eq.~\eqref{Model} by considering the classical limit $\Omega = J_\perp = 0$. The ground states of the resulting classical model were determined by Bak and Bruinsma\cite{Bak82} who demonstrated that the phase diagram as a function of $\Delta$ possesses infinitely many phases, and the magnetization $m(\Delta)$ forms a complete devil's staircase. For $\Delta < 0$ the ground state is fully polarized $...\downarrow \downarrow \downarrow \downarrow...$ with magnetization $m = - \frac{1}{2}$ per site. For positive $\Delta$, it becomes energetically favorable to accommodate up-spins, i.e., Rydberg excitations in the ground state. The up-spins are however arranged in a way so that their distance is maximized thus minimizing their interaction energy. The resulting mean distance between two Rydberg excitations, known as the Rydberg blockade radius, depends on the density of excitations. By increasing $\Delta$, more Rydberg excitations are accomodated, the blockade radius shrinks and the magnetization increases in steps giving rise to a plethora of phases that are commensurate with the periodicity of the optical lattice potential.
The magnetization finally vanishes at sufficiently large detuning $\Delta$ for the antiferromagnetic phase $... \downarrow \uparrow \downarrow \uparrow\downarrow \uparrow ...$, where every second atom is excited. Generally, the commensurate phases break a certain $Z_p$ symmetry of the optical lattice where $p$ is the periodicity of the spin configuration. For example, the antiferromagnetic state repeats itself after every second site so that it breaks an Ising, i.e., a $Z_2$ symmetry. The devilish aspect of the phase diagram is due to its fractal appearance as phases of all possible commensurabilities with the optical lattice are realized on the classical level.

The presence of a finite $\Omega$ and $J_\perp$ induce quantum fluctuations, which eventually melt the commensurate phases,
and one enters the realm of commensurate-incommensurate transitions, floating phases and dislocation-mediated melting intensively studied at the dawn of the 1980's.\cite{ChaikinBook,Bak81} We are going to apply the results and techniques developed at that time to the present context. In particular, throughout this work we apply intensively the strategy of Villain and Bak\cite{Villain81} as used in their analysis of the axial next-nearest-neighbor Ising (ANNNI) model. First, we consider the classical Hamiltonian close to one of its phase transitions and analyze the effect of a finite hopping $J_\perp$ perturbatively still keeping $\Omega=0$. This allows to derive an effective low-energy Hamiltonian for excitations of the Rydberg crystal -- domain walls -- that represents a proper fixed-point theory in the renormalization group sense. We are then able to apply a standard stability analysis of this effective theory with respect to quantum fluctuations induced by a finite Rabi frequency $\Omega$, that creates topological defects, i.e., dislocations. This strategy allows us to derive the phase diagram in the limit $\Omega \ll J_\perp$. We then combine this knowledge with known results in order to derive the topology of the full phase diagram of the Hamiltonian Eq.~\eqref{Model}, in particular, in the experimentally relevant regime $\Omega \gg J_\perp$.

The rest of the article is organized as follows.
In section \ref{sec:CommPhases} we consider the various gapped commensurate phases where the Rydberg crystal is commensurate with the optical lattice and construct an exact critical theory for the transitions between those crystalline states. In section \ref{sec:KT} we consider the Hamiltonian \eqref{Model} in the continuum limit and discuss the melting of the incommensurate floating phase. We conclude
with a summary and discussion in section \ref{sec:summary}.
In appendix \ref{app:BEC} we shortly discuss the superfluid regime, which has been addressed in a recent experiment on Rydberg atoms in a 1D lattice.~\cite{Viteau}
In appendix \ref{app} we establish the effective particle formulation used in the text to describe arbitrary transitions between classical commensurate phases.

\section{Commensurate Rydberg-crystal phases}
\label{sec:CommPhases}

In this section, we derive the topology of the phase diagram of the Hamiltonian \eqref{Model} in the $(\Delta, \Omega)$ plane with an emphasis on the various Rydberg crystal phases that are commensurate with the optical lattice.

In order to approach the problem and to circumvent the complexity of the Hamiltonian \eqref{Model} at its classical level $\Omega = J_\perp = 0$, where it possesses a fractal phase diagram,\cite{Bak82} it is useful to consider first a class of auxiliary Hamiltonians of the same form as Eq.~\eqref{Model} but the infinite-range interaction, $J_{|r|}$, of Eq.~\eqref{LongRangeInt} replaced by an interaction $J^{(n)}_{|r|}$ of finite range only
\begin{align} \label{frInt}
\begin{array}{ll}
J^{(n)}_{|r|} > 0 &{\rm if}\quad 1 \leq |r| \leq n,
\\
J^{(n)}_{|r|} = 0 &{\rm if}\quad |r| > n.
\end{array}
\end{align}
This finite range interaction should have the property that it obeys the convexity condition, $J^{(n)}_{|r-1|}+J^{(n)}_{|r+1|}> 2 J^{(n)}_{|r|}$ for each $n$ and that it coincides with $J_{|r|}$ in the limit $n \to \infty$.
We start our analysis by considering only interactions between close neighbors with small $n$, and afterwards we discuss the modifications obtained by including interaction of higher order by increasing $n$. This allows us to develop the topology of the phase diagram step by step. The full Hamiltonian is obtained in the limit $n \to \infty$, but it turns out that for a finite hopping $J_\perp$ the topology of the phase diagram in the $(\Delta,\Omega)$ plane becomes insensitive to interaction components $J_{|r|}$ at
a sufficiently large distance $r$ so that $J_{|r|} \ll J_\perp$.

The case $n=1$ when the Rydberg interaction is limited to nearest-neighbors only, $J^{(1)}_{|r|} = J_R \delta_{|r|,1}$, is special and well-known, so we summarize here only the result. In this case, the Hamiltonian reduces to a standard strongly anisotropic spin-chain in a strong longitudinal and a weak transversal magnetic field. In the classical limit $\Omega = J_\perp = 0$, there are just two ground states, the antiferromagnetic state $...\uparrow \downarrow \uparrow \downarrow \uparrow...$ for $\Delta > 0$ and the fully polarized state  $...\downarrow\downarrow\downarrow \downarrow \downarrow ...$ for $\Delta < 0$. At the transition $\Delta = 0$, the classical Hamiltonian
reduces to Eq.~\eqref{HamCr} with
a macroscopically degenerate ground state manifold.
The corresponding zero-point entropy,
however, is immediately quenched by an infinitesimal hopping $J_\perp$ and the single transition is replaced by an extended Luttinger liquid phase bounded by two Lifshitz transitions.\cite{Trippe09} The Luttinger liquid phase, however, is itself unstable with respect to a finite transverse field $\Omega$, and it gives way to a single line of Ising transitions in the $(\Omega,\Delta)$ plane.\cite{Ovchinnikov03}

This picture, however, changes qualitatively upon including Rydberg interactions beyond nearest neighbors. In the next section \ref{sec:J1J2chain}, we consider the spin-chain with nearest and next-nearest neighbor interactions before discussing the long-range interaction in section \ref{sec:LRIntChain}.

\subsection{Anisotropic spin-chain with nearest and next-nearest neighbor interaction}
\label{sec:J1J2chain}

The auxiliary Hamiltonian with the infinite range interaction replaced by $J^{(2)}_{|r|}$ of Eq.~\eqref{frInt}, i.e., with nearest, $J_1$, and next-nearest neighbor interaction, $J_2$, reads
\begin{align} \label{J1J2Ham1}
\mathcal H =& \sum_{j} \sum_{r = 1,2} J_{r} \, \big(S_j^z+1/2\big) \big(S_{j+r}^z+1/2 \big)
\\\nn&
-\frac{J_\perp}{2} \sum_j \big( S^+_j S^-_{j+1} + \text{h.c.} \big)
-\Delta \sum_j S_j^z + \Omega \sum_j S_j^x.
\end{align}
It will be important that the interactions obey the condition of convexity $J_1 > 2 J_2 > 0$; see Eq.~\eqref{LongRangeInt}. In this section, we discuss in detail the phase diagram of Eq.~\eqref{J1J2Ham1} in the limit $\Omega, |\Delta|, J_\perp \ll J_1$. We start with a discussion of the classical ground states at $\Omega = J_\perp = 0$ and identify two phase transitions. We then analyze how the phase transitions are influenced by a finite $J_\perp$ and $\Omega$ and present the resulting phase diagram.

The discussion of this section is closely related to work by Fendley, Sengupta and Sachdev.\cite{Fendley04} In fact, the Hamiltonian \eqref{J1J2Ham1} in the limit $J_\perp \to 0$ and $J_1 \to \infty$ reduces to the one studied in Ref.~\onlinecite{Fendley04}. However, the presence of a finite $J_\perp$ in our case facilitates some of the analysis and allows to present the physics in a transparent manner.

\subsubsection{Classical analysis for $\Omega = J_\perp = 0$}
\label{subsec:classJ1J2}
At $\Omega =J_\perp=0$, there are three classical ground states as a function of the  longitudinal field $\Delta$ that are
separated by two classical phase transitions located at
\begin{align}
\Delta_{c1} = 0, \quad \Delta_{c2} = 3 J_2.
\end{align}
The classical ground states are the fully polarized state $..\downarrow \downarrow \downarrow \downarrow \downarrow..$ at $\Delta < \Delta_{c1}$ with magnetization $m=-1/2$, the state $..\downarrow \downarrow \uparrow \downarrow \downarrow  \uparrow \downarrow \downarrow \uparrow..$ with periodicity $p=3$ and magnetization $m=-1/6$ for intermediate fields $\Delta_{c1} < \Delta < \Delta_{c2}$, and the antiferromagnet $..\downarrow \uparrow \downarrow \uparrow \downarrow \uparrow..$  with periodicity $p=2$ and magnetization $m=0$ for $\Delta > \Delta_{c2}$. The two states with period $p$ break a $Z_p$ symmetry of the optical lattice; the state with $m=-1/6$ is a $Z_3$ symmetry-broken state and the antiferromagnetic state breaks an Ising $Z_2$ symmetry.

At both transitions, there is a finite residual entropy at zero temperature. For example, the lowest energy excitation of the antiferromagnet is a single pair of two adjacent down spins, i.e., a domain wall or a spinon $..\uparrow \downarrow \uparrow \downarrow\downarrow \uparrow \downarrow \uparrow..$. The excitation energy of these domain walls vanishes at the second critical field $\Delta_{c2}$ so that they proliferate, resulting in a large degeneracy of the ground state at this critical point. The domain walls are strongly interacting as they must be separated at least by a single down spin. The corresponding entropy per site can be evaluated by the transfer matrix method that gives in the thermodynamic limit $L \to \infty$
\begin{align} \label{ResEntropy1}
S = \frac{1}{2} \log x = 0.28119...\quad {\rm at}\quad \Delta=\Delta_{c2},
\end{align}
where $x$ is the real root of the cubic equation $-1+x-2 x^2 + x^3 = 0$. Similarly, the lowest energy excitation of the fully polarized state is a single up spin $..\downarrow \downarrow \downarrow \uparrow \downarrow \downarrow \downarrow..$ whose excitation energy vanishes at $\Delta_{c1} = 0$. At the critical field $\Delta_{c1}$ all states with up spins separated at least by two down spins are degenerate leading to a residual entropy
\begin{align}\label{ResEntropy2}
S = \frac{1}{2} \log x = 0.38224...\quad {\rm at}\quad \Delta=\Delta_{c1} = 0,
\end{align}
where $x$ is the real root of the equation $-1-2 x- x^2 + x^3 = 0$.

\begin{figure}
\includegraphics[width=0.4\textwidth]{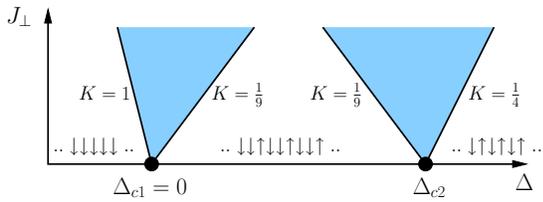}
\caption{Phase diagram in the $(\Delta,J_\perp)$ plane of the auxiliary Hamiltonian \eqref{J1J2Ham1} at $\Omega = 0$. A finite $J_\perp$ gives rise to Luttinger liquid phases (shaded areas) emanating from the classical transition points at $\Delta_{c1} = 0$ and $\Delta_{c2} = 3J_2$. The spin configurations indicate the classical ground states in the three phases with magnetizations $m = -1/2$, $m=-1/6$ and $m=0$ per site, respectively. The values of the Luttinger parameter $K$ at the four Lifshitz transitions given in the figure is derived in the text.
}
\label{fig:PDJphz}
\end{figure}

In the presence of a finite $J_\perp$ the excitations acquire kinetic energy and each single classical
transition is replaced by two C-IC transitions enclosing an extended floating Luttinger liquid phase; see Fig.~\ref{fig:PDJphz}. We first discuss these transitions close to the second critical field $\Delta_{c2}$ and afterwards the ones close to $\Delta_{c1}$.

\subsubsection{Transition at $\Delta_{c2}$}
\label{subsec:trans1}

We first focus on the transition at $\Delta_{c2}$ and study the influence of a finite hopping $J_\perp$ that leads to an extended Luttinger liquid phase. Afterwards, the stability of these phases is analyzed with respect to a finite driving $\Omega$.

\paragraph{effective Hamiltonian for small $J_\perp >0$ and $\Omega = 0$.}
\label{sec:StabLuttHc1}
At the classical transition located at $\Delta_{c2}$ the ground state changes its period from $p = 2$ to  $p = 3$.
Generally, the unit cell of a commensurate state with periodicity $p$ involves $p$ spins that we denote as $u_p$, for example $u_p = \downarrow \uparrow$ for $p=2$ and $u_{p'} = \downarrow \downarrow \uparrow$ for $p'=3$. Near the classical transition between two adjacent commensurate phases we can limit ourselves to the degenerate states with the successive occurrence of unit cells of the $p=2$ and $p'=3$ states only, e.g. $.. u_2 u_2 u_{3} u_2 u_{3} u_{3} ...$ The border between two different unit cells corresponds to a domain wall. In order to capture the meandering of domain walls we follow Villain and Bak\cite{Villain81,Rujan81} and introduce fictitious particles living on a fictitious lattice to describe the commensurability transition. We associate the unit cell $u_2$ with an empty site and the unit cell $u_{3}$ with an occupied site so that, e.g., the above sequence of unit cells is identified with $...001011...$ The number of particles is denoted as $N$ and the number of empty sites by $N_e$ so that $L = 2 N_e + 3 N$ where $L$ is the length of the chain. The number of fictitious particles is bounded by $0 \leq N \leq L/3$ and the length of the fictitious lattice is $L_f = N_e + N = (L-N)/2$. The magnetization per site is a function of $N$ and given by $m =  -N/(2L)$ ranging between $-1/6 \leq m \leq 0$.

At the classical transition, all states with $0 \leq N \leq L/3$ are degenerate. We can evaluate the resulting entropy $S = \frac{1}{L} \log \Omega_s$ where $\Omega_s$ is the total number of available states,
\begin{align}
\Omega_s = \sum^{L/3}_{N=0} \frac{\left(\frac{L-N}{2}\right)!}{N! \left(\frac{L-N}{2}-N\right)!},
\end{align}
where each term in the sum is just the number of arrangements of $N$ particles on $L_f=(L-N)/2$ sites. This exactly reproduces Eq.~\eqref{ResEntropy1} in the limit $L \to \infty$. Note that we are ignoring here boundary effects that depend on the employed boundary conditions and are negligible in the thermodynamic limit.

A finite $J_\perp$ allows the particles to propagate on the fictitious lattice. The effective Hamiltonian describing this propagation of domain walls is, in lowest order in $J_\perp$,
just the tight-binding model for hard-core bosons
\begin{align} \label{HeffHc1}
\mathcal{H}_{\rm eff} = \sum^{L_f}_{i=1}\left[
-\frac{J_\perp}{2} \left(a^\dagger_i a^\pdag_{i+1} + a^\dagger_{i+1} a^\pdag_i\right)
+ \frac{\Delta - \Delta_{c2}}{2} a^\dagger_i a^\pdag_{i}
\right],
\end{align}
with the hopping amplitude $J_\perp/2$. All states must obey the hard-core constraint $a^\dagger_i a^\pdag_i \leq 1$.
The effective chemical potential is obtained by realizing that each particle carries the magnetization $-\frac{1}{2}$ and, in addition, that at the critical field $\Delta = \Delta_{c2}$ for $J_\perp=0$ the energy is independent of the number of particles $N$.

The ground state energy in the presence of $N$ particles is easily evaluated,
\begin{align}
E = N \frac{\Delta - \Delta_{c2}}{2} - J_\perp \sum^{N/2}_{n=-N/2} \cos\left(\frac{2\pi}{L_f} n\right).
\end{align}
Using that $N = - 2 L m$ one obtains the magnetic energy per site, $\varepsilon = E/L$, as a function of $m$ that reads in the limit $L \to \infty$
\begin{align}
\varepsilon(m) =&
- m(\Delta - \Delta_{c2}) + \frac{J_\perp}{2\pi} (1+2m)\sin \left( \frac{4 \pi m}{1+2m} \right).
\end{align}
Minimizing $\varepsilon(m)$ with respect to the magnetization $-1/6 \leq m \leq 0$, one obtains the magnetic equation of state $\partial \varepsilon/\partial m = 0$. From the second derivative, one gets the differential susceptibility $\chi^{-1} = \partial^2 \varepsilon/\partial m^2$,
\begin{align} \label{SuscHc1}
\chi = \frac{(1+2m)^3}{8 \pi J_\perp \sin \left( \frac{-4 \pi m}{1+2m} \right) }.
\end{align}
The classical jump of the magnetization from zero magnetization to $m=-1/6$, implying an infinite susceptibility at the classical transition, is smoothened out by the finite hopping $J_\perp$. Instead, the classical transition is replaced by two Lifshitz transitions at the fields $\Delta = \Delta^\pm_{c2}$ with
\begin{align}
\Delta^+_{c2}(\Omega=0) &= \Delta_{c2} + 2 J_\perp,
\\
\Delta^-_{c2}(\Omega=0) &= \Delta_{c2} - 3 J_\perp.
\end{align}
Close to these critical fields the magnetization changes with the square-root cusps typical for C-IC transitions,\cite{Japaridze78,Pokrovsky79}
\begin{align}
m &= - \frac{1}{4\pi} \sqrt{\frac{\Delta^+_{c2} - \Delta}{J_\perp}} \Theta(\Delta^+_{c2}-\Delta) + \mathcal{O}\left(\frac{\Delta^+_{c2} - \Delta}{J_\perp}\right),
\\
m &= - \frac{1}{6} + \frac{1}{9\pi} \sqrt{
\frac{2(\Delta - \Delta^-_{c2})}{3 J_\perp}}
\Theta(\Delta - \Delta^-_{c2})
+ \mathcal{O}\left(\frac{\Delta - \Delta^-_{c2}}{J_\perp}\right).
\end{align}
For fields $\Delta^-_{c2} < \Delta < \Delta^+_{c2}$, the gas of domain walls is compressible and forms a Luttinger liquid, whose properties we consider in the following.

\paragraph{Luttinger liquid theory for $J_\perp>0$ and its stability with respect to $\Omega$ perturbations.}
\label{sec:StabLuttHc1}
The low-energy properties of the effective Hamiltonian \eqref{HeffHc1} in the magnetic field range $\Delta^-_{c2} < \Delta < \Delta^+_{c2}$ is captured by the Luttinger liquid theory
\be \label{LuttModel}
\mathcal{H}_{\rm LL} =\frac{v}{2\pi} \int dx \left( K (\partial_x \theta)^2+\frac{1}{K}(\partial_x \phi)^2 \right),
\ee
where the bosonic field $\phi$ represents fluctuations in the magnetization, $\delta m \sim - \frac{1}{\pi} \partial_x \phi$, and $[ \frac{1}{\pi}\partial_x \phi(x),\theta(x')]=- i\delta(x-x')$; the two parameters $v$ and $K$ are to be determined. The velocity $v$ is given by the energy spacing for a finite size system $\Delta_L = \frac{2 \pi}{L} v$. From the Hamiltonian \eqref{HeffHc1}, we get
$\Delta_L = \frac{2\pi J_\perp}{L_f} \sin(N \pi/L_f)$
and after expressing $N$ and $L_f$ in terms of the magnetization we obtain
\be \label{VelocityHc1}
v = \frac{2 J_\perp}{1+ 2 m} \sin \frac{- 4\pi m}{1 + 2 m}.
\ee
From the magnetic susceptibility \eqref{SuscHc1} and the relation\cite{GiamarchiBook} $\chi = K/(\pi v)$ we obtain the important result for the Luttinger parameter
\begin{align} \label{KHc1}
K = \left(m + \frac{1}{2} \right)^2.
\end{align}
As the upper boundary $\Delta^+_{c2}$  of the Luttinger liquid region is approached, the magnetization vanishes and the Luttinger parameter assumes the value $K=1/4$. Close to the lower boundary $\Delta^-_{c2}$ the magnetization becomes $m=-1/6$ and thus $K=1/9$; see also Fig.~\ref{fig:PDJphz}.
The values of the Luttinger parameter close to $\Delta^\pm_{c2}$ are consistent with the general expectations for Lifshitz transitions into commensurate states of periodicity $p$ that are stabilized by Umklapp scattering processes of $p$ spinons yielding $K=1/p^2$.\cite{GiamarchiBook}

Physically, the Luttinger liquid phase corresponds to an incommensurate, floating crystalline state of Rydberg excitations. It has no true long range order but algebraically decaying crystalline correlations\cite{GiamarchiBook}
\begin{equation} \label{FCcorrel}
\langle S^z(x) S^z(0) \rangle - m^2  \sim   \frac{\cos\left[ \pi (1+2 m) x \right]}{|x|^{2 K}}  \ .
\end{equation}

Now we are in a position to analyze the effect of the transversal magnetic field $\Omega$. Within the Luttinger liquid framework, the transversal field gives rise to the perturbation\cite{GiamarchiBook}
\begin{align} \label{KLHc1}
\mathcal{H}_{\rm pert} = \frac{\Omega}{\sqrt{2\pi} a} \int dx  \cos(\theta),
\end{align}
where $a$ is the optical lattice spacing. Note that the interaction $J_\perp$ is ferromagnetic so that the low-energy fluctuations of the $S^x$ operator are non-staggered.
The origin of this perturbation becomes apparent by realizing that the conjugate fields $\phi(x)$ and $\theta(x)$ can be viewed as the polar- and azimuthal angles of classical vectors representing the spins. The transverse field term \eqref{KLHc1} thus wants to lock the azimuthal angles at $\theta=0$ or $\pi$, depending on the sign of $\Omega$, such that all spins point in x-direction, as expected.
Using standard scaling arguments,\cite{GiamarchiBook} one finds that the perturbation \eqref{KLHc1} is relevant in the renormalization group (RG) sense if the Lutttinger parameter exceeds a critical value $K \geq 1/8$. Using Eq.~\eqref{KHc1} this implies for the magnetization $m \leq -(2-\sqrt{2})/4$ and finally $\Delta \geq \Delta_{c2}^{\rm KT}$ where the critical field is determined by the magnetic equation of state
\begin{align} \label{KT-Hc1}
\Delta^{\rm KT}_{c2} &=  \Delta_{c2} - J_\perp \left( \frac{\sin(2\sqrt{2} \pi)}{\pi} - 2 \sqrt{2} \cos(2\sqrt{2} \pi) \right)
\\\nn&
= \Delta_{c2} - J_\perp \times 2.5908...
\end{align}
So the Luttinger liquid generated by $J_\perp$ is only stable very close to the first Lifshitz transition at $\Delta^-_{c2}$ in a small field range obeying $- 3 <  (\Delta - \Delta_{c2})/J_\perp < - 2.5908..$. At the critical field $\Delta^{\rm KT}_{c2}$ the operator \eqref{KLHc1} leads to a Kosterlitz-Thouless transition into an incommensurate gapped state. This gapped phase extend to finite $\Omega$, and the phase boundary in the $(\Delta, \Omega)$ plane starts at $\Delta^{\rm KT}_{c2}$ linearly with $\Omega$, which follows
from the separatrix of the Kosterlitz-Thouless RG flow.\cite{Kosterlitz73}

\paragraph{Effective Hamiltonian for small finite $J_\perp$ and $\Omega$.}
It is instructive to consider the modification of the effective model \eqref{HeffHc1} due to the transversal magnetic field explicitly with the help of a Schrieffer-Wolf transformation. Including corrections up to first order in $J_\perp$ and second order in $\Omega$, i.e., $\mathcal{O}(J_\perp, \Omega^2)$, the low-energy Hamiltonian assumes the form
\begin{align} \label{HeffHc1-2}
\lefteqn{\mathcal{H}_{\rm eff} =}
\\\nn& \sum^{L_f}_{i=1} \left[-t
\left(a^\dagger_i a^\pdag_{i+1} + a^\dagger_{i+1} a^\pdag_i\right)
- \mu a^\dagger_i a^\pdag_{i}
+ U a^\dagger_i a^\pdag_{i} a^\dagger_{i+1} a^\pdag_{i+1}
\right],
\end{align}
where we remind that the size of the fictitious lattice $L_f = N_e + N = (L-N)/2$ depends on the number of particles $N = \sum_i a^\dagger_i a^\pdag_{i}$. The hopping amplitude $t$ at $\Delta = \Delta_{c2}$ is given by
\begin{align}
t &= \frac{J_\perp}{2} + \left(\frac{\Omega}{2}\right)^2 \left( \frac{1}{J_1-2J_2} + \frac{1}{2J_2}
\right).
\end{align}
To obtain the chemical potential $\mu$ and the interaction $U$, it is useful to consider the correction to the energy per site, $\delta \varepsilon = \delta E/L$, of various specific states in order $\mathcal{O}(\Omega^2, \Delta - \Delta_{c2})$ in the absence of any hopping $t$,
\begin{align}
\label{firstline}
\delta \varepsilon_{..0000..} &= - \left(\frac{\Omega}{2}\right)^2 \frac{1}{2} \left( \frac{1}{2J_1 - 3J_2}+\frac{1}{J_2} \right),
\\
\delta \varepsilon_{..1111..} &= \frac{\Delta - \Delta_{c2}}{6} - \left(\frac{\Omega}{2}\right)^2 \frac{1}{3} \left( \frac{2}{J_1 - 2 J_2}+\frac{1}{3 J_2} \right),
\\
\delta \varepsilon_{..010101..} &= \frac{\Delta - \Delta_{c2}}{10}
\\ &\quad\nn
- \left(\frac{\Omega}{2}\right)^2 \frac{1}{5} \left( \frac{1}{2J_1 - 3 J_2}+\frac{1}{J_2} + \frac{2}{J_1 - 2 J_2}\right).
\end{align}
Equation~(\ref{firstline}) gives the correction to the vacuum with no particles $..0000.. \equiv ..u_2 u_2 u_2 u_2..$ where $u_2 = \downarrow \uparrow$. The energy correction $\delta \varepsilon_{..010101..}$ for the state with particles at every second site can be used to compute the chemical potential; for $N_e = N$ which implies $N=L/5$ we have the relation $-\mu N = (\delta \varepsilon_{..010101..} - \delta \varepsilon_{..0000..}) L$. With this we obtain for the chemical potential
\begin{align}
\mu &= \frac{\Delta_{c2} - \Delta}{2} + 5\left(\delta \varepsilon_{..0000..} - \delta \varepsilon_{..010101..}\right)
\\\nn
&=
\frac{\Delta_{c2} - \Delta}{2} - \frac{\Omega^2}{8}\left(
\frac{3}{2J_1 - 3J_2} -\frac{4}{J_1-2J_2} +\frac{3}{J_2}
\right).
\end{align}
The interaction $U$ can be obtained by relating the energy corrections to the vacuum and to the state with particles at every site,
 $\delta \varepsilon_{..1111..} L = \delta \varepsilon_{..00000..}L +N( - \mu +U)$ where the number of particles is now maximal, i.e., $N=L/3$. This yields an attractive interaction given by
 \begin{align}
 U &= \mu + 3 (\delta \varepsilon_{..1111..} -\delta \varepsilon_{..0000..} )
 = - \left(\frac{\Omega}{2}\right)^2 \frac{1}{3 J_2}.
 \end{align}

Alternatively, we can consider the effective Hamiltonian after a particle-hole transformation where holes are created by $h_i^\dagger$ within the fully occupied chain $..111111..$,
\begin{align} \label{HeffHc1-holes}
\lefteqn{\mathcal{H}_{\rm eff} = }\\\nn
&\sum^{L_f}_{i=1} \left[-t_e
\left(h^\dagger_i h^\pdag_{i+1} + h^\dagger_{i+1} h^\pdag_i\right)
- \mu_e h^\dagger_i h^\pdag_{i}
+ U_e h^\dagger_i h^\pdag_{i} h^\dagger_{i+1} h^\pdag_{i+1}
\right].
\end{align}
Again, all states must obey the hard-core constraint $h^\dagger_i h^\pdag_i \leq 1$.
The hopping amplitude is the same as for the particles $t_e = t$.
However, the effective chemical potential, $\mu_e$, and the interaction, $U_e$, for the holes are now obtained from the relations $-\mu_e N_e = (\delta \varepsilon_{..010101..} - \delta \varepsilon_{..1111..}) 5 N_e$ and $\delta \varepsilon_{..0000..} 2 N_e = \delta \varepsilon_{..11111..} 2 N_e +N_e( - \mu_e +U_e)$, respectively. In particular, this yields the relation $\mu_e = -2 \mu/3 +5U/3$.

At this order of perturbation theory, the model \eqref{HeffHc1-2} still exhibits two Lifshitz transitions. To obtain the upper critical field we can neglect the interaction and simply set $2t + \mu = 0$, yielding
\begin{align}\label{Delta+c2}
\Delta^+_{c2}(\Omega) &= \Delta_{c2} + 2 J_\perp
\\\nn&\quad
+ \left(\frac{\Omega}{2}\right)^2\left(
-\frac{3}{2J_1 -3J_2} + \frac{8}{J_1 - 2J_2}  - \frac{1}{J_2}
\right).
\end{align}
The Lifshitz transition at the lower  critical field takes place if $2 t_e + \mu_e = 0$, giving
\begin{align} \label{Delta-c2}
\Delta^-_{c2}(\Omega) &= \Delta_{c2} - 3 J_\perp
\\\nn&\quad
- \left(\frac{\Omega}{2}\right)^2\left(
\frac{3}{2J_1 -3J_2} + \frac{2}{J_1 - 2 J_2} +\frac{13}{3 J_2}
\right).
\end{align}

However, the picture changes qualitatively if corrections of order $\mathcal{O}(J_\perp \Omega)$ or of order $\mathcal{O}(\Omega^3)$ are included as processes are generated in the effective description that do not conserve the number of particles. For example, applying three spin flips two effective particles can be created out of three adjacent holes, so that the state $..\downarrow \uparrow \downarrow \uparrow\downarrow \uparrow..$ = $..000..$ is converted to $..\downarrow \downarrow \uparrow \downarrow\downarrow \uparrow..$ = $..11..$. In the dense limit close to $\Delta_{c2}^-$ where there are only few holes in the ground-state, this process corresponds to the additional operator $h_i h_{i+1} h_{i+2}$ in the Hamiltonian that destroys three holes. This operator is irrelevant and the Lifshitz transition near $\Delta_{c2}^-$ remains unaffected. In the dilute limit, however, this process corresponds to the additional operator $a^\dag_i a^\dag_{i+1}$ that creates two particles which is relevant and converts the Lifshitz transition at $\Delta_{c2}^+$ into an Ising transition.
This is analogous to the mechanism at play close to the multicritical point in the ANNNI model where also an Ising transition emerges.\cite{Villain81} As a result, there is also an additional transition of Kosterlitz-Thouless type at $\Delta_{c2}^{\rm KT}$, with $\Delta_{c2}^- < \Delta_{c2}^{\rm KT} < \Delta_{c2}^+$,  between a gapped ground state and the Luttinger liquid as discussed in Section \ref{sec:StabLuttHc1}. In order to evaluate the correction to $\Delta^{\rm KT}_{c2}$ in second order in $\Omega$, however, the interacting Hamiltonian \eqref{HeffHc1-2} must be solved.\cite{Fendley04}

\subsubsection{Transition at $\Delta_{c1}$}
\label{subsec:trans2}

We now turn to the transition at $\Delta_{c1}$ between the fully polarized state and the commensurate phase with magnetization $m = - \frac{1}{6}$; see Fig.~\ref{fig:PDJphz}. Again, we first study the influence of a finite hopping $J_\perp$ and afterwards we perform a stability analysis with respect to $\Omega$ perturbations.

\paragraph{Effective Hamiltonian for small $J_\perp >0$ and $\Omega = 0$.}
We can repeat the reasoning of the previous section and derive an effective Hamiltonian also close to the transition at $\Delta_{c1}=0$. Whereas the ground state for $\Delta_{c1} < \Delta < \Delta_{c2}$ has periodicity $p=3$ with unit cell $u_3 = \downarrow \downarrow\uparrow$, the ground-state for $\Delta < 0$ is fully polarized. The unit cell of the latter is trivial as it consists of a single spin only, $u_1 = \downarrow$. At the classical transition $\Delta = 0$ for $J_\perp = \Omega = 0$, there is again a class of degenerate ground states that correspond to the successive occurrence of the two unit cells $u_3$ and $u_1$, e.g., $..u_3 u_3 u_1 u_3 u_1 u_1 ..$ We introduce effective particles and associate $u_1$ with an empty and $u_3$ with an occupied site so that the above sequence becomes $..110100..$. The length of the chain is related to the number of particles, $N$, with $0 \le N \le L/3$, and the number of empty sites $N_e$ by $L = N_e + 3 N$. The particles live on a fictitious lattice of size $L_f = N_e + N = L-2N$ and the magnetization per site is given by $m = -(N_e+N)/(2L) = N/L - 1/2$ that varies between $-1/2 \leq m \leq -1/6$.

In the classical limit, there is again a macroscopic degeneracy at the critical field $\Delta_{c1}$ with a finite entropy per site $S = \frac{1}{L}\log \Omega_s$,
\begin{align}
\Omega_s = \sum_{N=0}^{L/3} \frac{(L-2 N)!}{N! ( L-2 N)!},
\end{align}
where each term corresponds to the number of arrangement of $N$ particles on $L_f$ sites. This reproduces the entropy of Eq.~\eqref{ResEntropy2}.

A finite $J_\perp$ results in hopping of particles described by the tight-binding model of hard-core bosons,
\begin{align} \label{HeffHc2}
\mathcal{H}_{\rm eff} = \sum^{L_f}_{i=1}\left[
-\frac{J_\perp}{2} \left(a^\dagger_i a^\pdag_{i+1} + a^\dagger_{i+1} a^\pdag_i\right)
- (\Delta - \Delta_{c1}) a^\dagger_i a^\pdag_{i}
\right],
\end{align}
where the chemical potential is obtained with the same reasoning as for Eq.~\eqref{HeffHc1}.
Repeating the steps of the previous analysis in section \ref{subsec:trans1}, we obtain for the magnetic energy per site
\begin{align}
\varepsilon(m) =  -m (\Delta - \Delta_{c1}) - \frac{2 J_\perp}{\pi} \cos\left(\frac{\pi}{4 m}\right).
\end{align}
Minimization with respect to $m$ gives the magnetic equation of state, $\partial \varepsilon/\partial m = 0$, that relates $\Delta$ to the magnetization $m$ in the range $-1/2 < m < -1/6$. For the susceptibility $\chi^{-1} = \partial^2 \varepsilon/\partial m^2$, we obtain
\begin{align}\label{SuscHc2}
\chi = \frac{8 m^3}{J_\perp \pi \cos\left(\frac{\pi}{4 m}\right)}.
\end{align}
The hopping $J_\perp$ smoothens out the classical transition and gives rise to two Lifshitz transitions located at
\begin{align}
\Delta^-_{c1}(\Omega=0) = -J_\perp, \quad {\rm and}\quad
\Delta^+_{c1}(\Omega=0) = 3 J_\perp,
\end{align}
close to which the magnetization again varies as a square root of $\Delta - \Delta^-_{c1}$ and $\Delta^+_{c1}-\Delta $, respectively.

\paragraph{Luttinger liquid theory for $J_\perp>0$ and its stability with respect to $\Omega$ perturbations.}
Between the critical fields $\Delta^-_{c1} < \Delta < \Delta^+_{c1}$, the magnetization is compressible and its fluctuations are described by a Luttinger liquid; see Eq.~\eqref{LuttModel}. In the following, we determine the velocity $v$ and Luttinger parameter $K$. The velocity is again obtained from the finite size spectrum of Eq.~\eqref{HeffHc2}, $\Delta_L = 2\pi v/L$, giving
\begin{align}
v = \frac{J_\perp }{2m}\cos\left(\frac{\pi}{4 m}\right).
\end{align}
With the help of the relation for the susceptibility\cite{GiamarchiBook} $\chi = K/(\pi v)$ we obtain the important result for the Luttinger parameter
\begin{align} \label{KHc2}
K = (2 m)^2.
\end{align}
The Luttinger parameter varies within the compressible regime between $K = 1/9$ close to the upper critical field $\Delta^+_{c1}$ where $m = -1/6$, and $K=1$ close to $\Delta^-_{c1}$ where the magnetization is saturated, $m=-1/2$; see Fig.~\ref{fig:PDJphz}.

A finite transversal field $\Omega$ gives rise to the operator \eqref{KLHc1} that becomes relevant for $K \geq 1/8$ implying for the magnetization the bound $m \geq -1/\sqrt{32}$ for the stability of the Luttinger liquid.
Using the equation of state, we obtain the corresponding bound for the field $\Delta \geq\Delta^{\rm KT}_{c1}$ with
\begin{align} \label{KTHc2}
\Delta^{\rm KT}_{c1} &= - \frac{2J_\perp}{\pi}  \left(\cos(\sqrt{2} \pi) + \sqrt{2} \pi \sin(\sqrt{2} \pi)  \right)
\\\nn&=J_\perp\times
 2.89583...
 \end{align}
 The Luttinger liquid is only stable within a very narrow range $2.8958.. < \Delta/J_\perp < 3$.
At the critical field $\Delta^{\rm KT}_{c1}$ the operator \eqref{KLHc1} results in a Kosterlitz-Thouless transition destroying the Luttinger liquid.

\paragraph{Effective Hamiltonian close to $\Delta^+_{c1}$ for small $\Omega, J_\perp > 0$.}
In the following, we discuss correction to the effective Hamiltonian \eqref{HeffHc2} that are generated by a finite transversal field $\Omega$. In the presence of $\Omega$, the number of fictitious particles is not conserved anymore already in linear order in $\Omega$. In the dilute limit close to $\Delta_{c1}^{-}$, the transversal field $\Omega$ can create particles by flipping a single spin only, i.e., it can produce $u_3$ unit cells within the fully polarized state $..\downarrow \downarrow\downarrow\downarrow.. = ..u_1 u_1 u_1 u_1..$. In first order in $\Omega$, an operator of the form $a^\dagger_i$ is generated in the Hamiltonian that is relevant and immediately destabilizes the Luttinger liquid in agreement with Eq.~\eqref{KTHc2}. In the dense limit close to $\Delta_{c1}^{+}$, on the other hand, the transversal field can convert in linear order in $\Omega$ a particle, i.e., a $u_3 = \downarrow\downarrow \uparrow$ unit cell by down-flipping the last spin into three holes. This process corresponds to an additional term in the Hamiltonian that consists of the product of three hole-creation operators $h^\dagger_i h^\dagger_{i+1} h^\dagger_{i+2}$. This operator is irrelevant in the limit of small densities of holes,
and the Luttinger liquid survives in a finite sliver close to $\Delta^+_{c1}$.

This actually allows to determine the $\Omega$-dependence of the phase boundary $\Delta^+_{c1}$ in the limit $\Omega \to 0$. It can be determined by solving the Schr\"odinger equation for a single hole in the background of particles in lowest order in $\Omega$. The excitation energy of the hole vanishes at $\Delta^+_{c1}$. Equivalently, we can consider the
effective single hole Hamiltonian that captures the Lifshitz transition
\begin{align}
\mathcal{H}^{\rm cr}_{\rm eff} &= \sum^{L_f}_{i=1} \left[
-t_e
\left(h^\dagger_i h^\pdag_{i+1} + h^\dagger_{i+1} h^\pdag_i\right)
- \mu_e h^\dagger_i h^\pdag_{i}
\right].
\end{align}
The three hole-creation operator generated in linear order in $\Omega$ and the interaction between holes can be neglected as they are irrelevant in the RG sense close to the Lifshitz transition. The hopping and the chemical potential in lowest order in $\Omega$ can be evaluated similar as in the previous section,
\begin{align}
t_e &= \frac{J_\perp}{2} + \left(\frac{\Omega}{2}\right)^2 \left(
\frac{1}{\Delta} + \frac{1}{J_1-\Delta}
\right),
\\
\mu_e &= -\frac{\Delta}{3} - \left(\frac{\Omega}{2}\right)^2 \Big(
\frac{1}{3\Delta} - \frac{2}{J_1 - \Delta}
\\&\nn\quad
- \frac{1}{2J_2-\Delta} + \frac{8}{3(J_1+J_2-\Delta)}
\Big).
\end{align}
Setting $2 t_e + \mu_e =0$ we obtain the critical field $\Delta^+_{c1}$ up to order $\mathcal{O}(J_\perp, \Omega^2/J_\perp)$
\begin{align} \label{Delta+c1}
\Delta^+_{c1} = 3 J_\perp + \frac{5\Omega^2}{12 J_\perp},
\end{align}
where we neglected corrections of order $\Omega^2/J_2$ and $\Omega^2/J_1$.

\subsubsection{Phase diagram in the $(\Delta, \Omega)$ plane}

\begin{figure}
\includegraphics[width=0.4\textwidth]{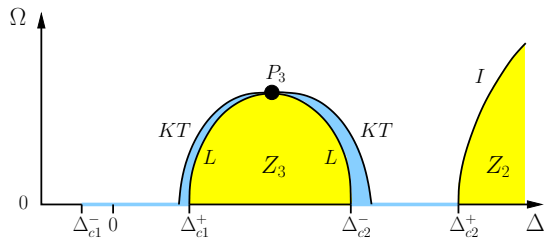}
\caption{Sketch of the phase diagram in the $(\Delta,\Omega)$ plane of the auxiliary Hamiltonian \eqref{J1J2Ham1} at a small but finite hopping $J_\perp$ with two symmetry broken phases (yellow shaded). A broken $Z_2$ phase is bounded by an Ising transition ($I$) and a broken $Z_3$ phase is enclosed by two Lifshitz ($L$), i.e., C-IC transitions  except at a single point where the transition is in the $Z_3$ Potts universality class ($P_3$). The Luttinger liquid phases (blue shaded) are confined to tiny regions around the $Z_3$ phase and are separated from the incommensurate gapped phase by a Kosterlitz-Thouless transition (KT).
}
\label{fig:PDhxhz}
\end{figure}

We started our analysis in section \ref{subsec:classJ1J2} with the classical limit of the Hamiltonian \eqref{J1J2Ham1}, $J_\perp = \Omega = 0$, where two transitions as a function of $\Delta$ were identified. A finite hopping $J_\perp$ replaces each of these two classical transitions with two extended gapless phases (see Fig.~\ref{fig:PDJphz}) where the fluctuations of the magnetization are described by the Luttinger liquid theory. As a consequence, at $\Omega = 0$ and $J_\perp > 0$ one crosses four Lifshitz transitions as a function of increasing $\Delta$. This picture is, however, modified if a finite $\Omega$ is taken into account as shown in Fig.~\ref{fig:PDhxhz}, which summarizes the phase diagram in the $(\Delta,\Omega)$ plane at some small but finite $J_\perp$.

The $\Delta$ axis in Fig.~\ref{fig:PDhxhz} corresponds to a horizontal cut through the phase diagram of Fig.~\ref{fig:PDJphz}. The Luttinger liquid phases (blue shaded) that are generated by the finite hopping $J_\perp$ and extend for $\Omega = 0$ between $\Delta^-_{ci} < \Delta < \Delta^+_{ci}$ with $i=1,2$, are now quenched by the finite driving $\Omega>0$ to only small regions close to $\Delta^+_{c1}$ and $\Delta^-_{c2}$. (The extension of the Luttinger phases is exaggerated in the figure in order to visualize them). Both small regions surround a commensurate phase with a broken $Z_3$ symmetry which is accompanied by a magnetization plateau at $m=-\frac{1}{6}$. The C-IC transition between this commensurate phase and the incommensurate Luttinger liquids is of Lifshitz-type, i.e., they are in the standard C-IC universality class\cite{Japaridze78,Pokrovsky79} so that the magnetization $m(\Delta)$ close to these transitions changes with the characteristic square root cusps. The values of the Luttinger parameter at these transitions is $K=\frac{1}{9}$; see Eqs.~\eqref{KHc1} and \eqref{KHc2}. We evaluated the phase boundaries $\Delta^+_{c1}(\Omega)$ and $\Delta^-_{c2}(\Omega)$ in lowest order in $\Omega$ and $J_\perp$; see Eqs.~\eqref{Delta-c2} and \eqref{Delta+c1}, respectively. The Luttinger liquids themselves are bounded by two Kosterlitz-Thouless (KT) transitions into gapped incommensurate phases. At these transitions, the Luttinger parameter reaches the value $K = \frac{1}{8}$ at which the operator $S^x_i$ driving the transitions becomes relevant. The positions $\Delta^{\rm KT}_{ci}(\Omega)$ of the two KT transitions, $i=1,2$, in the limit $\Omega \to 0$ were evaluated in Eqs.~\eqref{KT-Hc1} and \eqref{KTHc2}. From the RG flow of the KT transition it follows that the phase boundaries start linearly, $\Delta^{\rm KT}_{ci}(\Omega) - \Delta^{\rm KT}_{ci}(0) \sim \Omega$.

We did not calculate the phase diagram for larger values of $\Omega$. From previous works treating similar problems,\cite{Haldane83,Schulz83,Fendley04} the presence of a special point indicated by $P_3$ in Fig.~\ref{fig:PDhxhz} is known where a direct transition between the broken $Z_3$ phase and the gapped IC phase takes place which is in the $Z_3$ Potts universality class. We assumed here that the two Luttinger liquids meet at $P_3$.\cite{Abernathy93}
In the limit $J_1 \gg J_2$, the position of this quantum phase transition can be estimated to be at $\Omega \sim \Delta \sim J_2$.\cite{Fendley04}

Finally, at larger $\Delta$ the gapped IC phase is separated from the $Z_2$ broken phase by an Ising transition (I) whose boundary $\Delta^+_{c2}(\Omega)$ was evaluated in Eq.~\eqref{Delta+c2}.

We considered the phase diagram for a finite $J_\perp$ and evaluated the phase boundaries perturbatively in $\Omega$. In the limit $J_\perp \to 0$, the phase boundaries $\Delta^-_{c2}$, $\Delta^{\rm KT}_{c2}$, and $\Delta^+_{c2}$ will merge for $\Omega \to 0$; see Eqs.~\eqref{Delta-c2} and \eqref{Delta+c2}. It was demonstrated explicitly in Ref.~\onlinecite{Fendley04} that in the limit, $J_\perp \to 0$, the Luttinger liquid phase bounded by $\Delta^{-}_{c2}$ and $\Delta^{\rm KT}_{c2}$ nevertheless survives for $\Omega>0$.  Note that the evaluation of the phase boundary $\Delta^+_{c1}(\Omega)$ in Eq.~\eqref{Delta+c1} is only controlled if $\Omega \ll J_\perp$ and the limits $\Omega \to 0$ and $J_\perp \to 0$ do not commute.

\subsection{Anisotropic spin chain with long-range interactions}
\label{sec:LRIntChain}

In the last section, we demonstrated that the presence of an additional next-neighbor interaction gives rise to an additional broken $Z_3$ phase that is surrounded by a floating Luttinger liquid phase.  Now we consider the additional modifications caused by including longer range interactions. We can anticipate that an interaction between further distant neighbors stabilizes commensurate phases of higher periodicities $p$ breaking $Z_p$ symmetries of the lattice.

To prepare the basis for the following discussions, we first generalize the treatments of sections \ref{subsec:trans1} and \ref{subsec:trans2} to classical transition points between two arbitrary adjacent commensurate phases and study the influence of a finite hopping $J_\perp$. Afterwards we discuss the resulting modifications to the phase diagram.

\subsubsection{General transitions between commensurate phases at finite $J_\perp>0$ and $\Omega=0$}
\label{sec:pp'Transitions}

If a Hamiltonian with a finite-range interaction \eqref{frInt} is considered with $n>2$, higher order commensurate phases will be stabilized on the classical level. Here, following closely section \ref{sec:J1J2chain}, we consider the influence of a finite hopping $J_\perp$ on the classical transition between two arbitrary adjacent commensurate phases with periodicity $p$ and $p'$ and magnetization $m_p$ and $m_{p'}$, respectively.

We assume $m_p < m_{p'}$, and we will also use the relation between magnetizations and periodicities
\begin{align} \label{mm}
m_{p'} p p' = 1+ m_p p p'.
\end{align}
The latter can be derived from results of Bak and Bruinsma.\cite{Bak82} They demonstrated that
as a function of increasing $\Delta$ commensurate phases are successively stabilized such that their fraction $q/p$ of up-spins and, thus, their magnetization $m_p = q/p-1/2$ increases monotonously. It is elementary to show that for a given commensurate phase with a fraction $q/p$, the adjacent phase with larger fraction $q'/p'$ of up-spins can be found by solving the equation $q p'=  p q' - 1$, and choosing the solution with largest $p'$. The choice for $p'$ is only bounded by the range of the interaction, i.e., by $p' \le n+1$ with $n$ of Eq.~\eqref{frInt}. Substituting the magnetizations for $q$ and $q'$ one arrives at the above equation.

Near the classical transition we can limit ourselves to the degenerate states with the successive occurrence of unit cells of the $p$ and $p'$ states only, e.g. $.. u_p u_p u_{p'} u_p u_{p'} u_{p'} ...$. For a  definition of these unit cells corresponding to arbitrary fractions $q/p$ and $q'/p'$, and for a proof that the above limited set of states becomes the degenerate ground state manifold at the classical transitions; see appendix \ref{app}.
We again introduce fictitious particles living on a fictitious lattice and associate the unit cell $u_p$ with an empty site and the unit cell $u_{p'}$ with an occupied site so that the above sequence of unit cells is identified with $...001011...$ The number of particles is denoted as $N$ and the number of empty sites by $N_e$, so that $L = p N_e + p' N$ with the length of the chain $L$. The maximum number of fictitious particles is $0 \leq N \leq L/p'$ and the length of the fictitious lattice is $L_f = N_e + N = (L-N(p'-p))/p$. The magnetization per site is a function of $N$ and given by
$m =  m_p + N/(p L)$
ranging between $m_p \leq m \leq m_{p'}$.

At the classical transition, all states with $0 \leq N \leq L/p'$ are degenerate. We can evaluate the resulting entropy $S = \frac{1}{L} \log \Omega_s$ where $\Omega_s$ is the total number of available states,
\begin{align} \label{degeneracy}
\Omega_s = \sum^{L/p'}_{N=0} \frac{\left(\frac{L-N(p'-p)}{p}\right)!}{N! \left(\frac{L-N(p'-p)}{p}-N\right)!},
\end{align}
where each term in the sum is just the number of arrangements of $N$ particles on $L_f=(L-N(p'-p))/p$ sites.

A finite $J_\perp$ allows the particles to propagate on the fictitious lattice. Note that adjacent unit cells with different periodicity, $u_p$ and $u_{p'}$, can be exchanged by applying once the hopping operator with amplitude $J_\perp$; see appendix \ref{app}. So we obtain the effective Hamiltonian describing the propagation of fictitious particles in terms of a tight-binding model for hard-core bosons,
\begin{align}
\label{effgeneralized}
\mathcal{H}_{\rm eff} = \sum^{L_f}_{i=1}\left[ - \frac{J_\perp}{2}
 \left(a^\dagger_i a^\pdag_{i+1} + a^\dagger_{i+1} a^\pdag_i\right)
- \frac{\Delta - \Delta_c}{p}
a^\dagger_i a^\pdag_{i}
\right].
\end{align}
The chemical potential is obtained with the help of the relation between the number of particles $N$ and the magnetization $m$. The resulting magnetic energy per site reads
\begin{align}
\varepsilon(m) &= - m (\Delta - \Delta_c)
+
\frac{J_\perp}{\pi}
(m (p-p') + m_{p'} p' - m_p p)
\nn\\&\quad \times
\sin \left( \frac{\pi (m-m_{p'}) p'}{m (p-p') + m_{p'} p' - m_p p} \right),
\end{align}
that has to be minimized with respect to $m$, $\partial \varepsilon/\partial m = 0$ with $m_p \leq m \leq m_{p'}$. The finite $J_\perp$ gives rise to a compressible phase for fields $\Delta$ in the range
\begin{align} \label{LLExt}
\Delta_c - p J_\perp  < \Delta < \Delta_c + p' J_\perp.
\end{align}
The magnetization varies continuously within this range, and at both boundaries a Lifshitz transition takes place. As before we can evaluate the susceptibility $\chi$ and with the help of the velocity of the domain walls we derive the Luttinger parameter within the compressible phase
\begin{align}
K = \left((m_p - m) p - (m_{p'} - m ) p'  \right)^2.
\end{align}
With the help of the equation Eq.~\eqref{mm}, we obtain the value of the Luttinger parameter close to the Lifshitz transition into a state with periodicity $p$
\begin{align} \label{KLpp}
K_L = \frac{1}{p^2}.
\end{align}
This value validates that the commensurate phases are stabilized by Umklapp scattering of $p$ spinons.\cite{GiamarchiBook}

\subsubsection{Higher order commensurate phases for longer range interactions}
\label{sec:HigherOrderCP}

\begin{figure}
\includegraphics[width=0.4\textwidth]{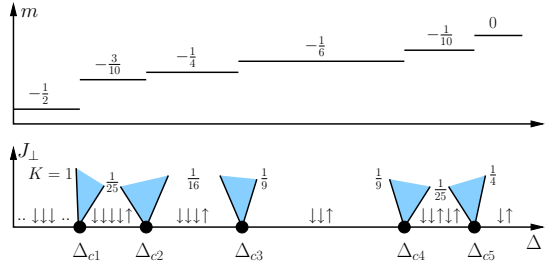}
\caption{Upper panel shows the magnetization as a function of $\Delta$ of the classical auxiliary Hamiltonian with a finite range interaction $J^{(4)}_{|r|}$ of Eq.~\eqref{frInt}. There are five classical transitions; see Eqs.~\eqref{J5CrDelta}.
The spin configurations in the lower panel illustrate the structure of the unit cells of the periodic commensurate spin states. A finite $J_\perp$ gives rise to Luttinger liquid phases (shaded regions) [see Eq.~\eqref{LLExt}]; the value of the parameter $K$ at their edges is given by Eq.~\eqref{KLpp} and shown in the plot. The figure is not to scale; the extension of phases with larger unit cells is exaggerated.
}
\label{fig:PDJ5-1}
\end{figure}

If a longer range interaction is considered, higher order commensurate phases appear on the classical level.\cite{Bak82} For example, with the interaction $J^{(n)}_{|r|}$ of Eq.~\eqref{frInt} with $n=4$, there are already six classical ground states at $J_\perp = \Omega =0$ in the vicinity of $\Delta = 0$. The positions of the phase transitions between those states are given by\cite{Bak82}
\begin{align} \label{J5CrDelta}
\begin{array}{ll}
\Delta_{c1} = 0, &
\Delta_{c2} = 5 J_4,
\\
\Delta_{c3} = 4 J_3 - 3 J_4, &
\Delta_{c4} = 3 J_2 - 2 J_3,\\
\Delta_{c5} = 3 J_2 - 2 J_3 + 5 J_4. &
\end{array}
\end{align}
The sequence of these classical ground states is illustrated in Fig.~\ref{fig:PDJ5-1}. The figure is not to scale; we exaggerated the extension of phases with larger periodicities in order to visualize them. For $\Delta < \Delta_{c1} = 0$, the fully polarized state is the ground state with magnetization $m = - \frac{1}{2}$. For positive $\Delta$ commensurate phases with various periodicities $p$ are stabilized with the concomitant breaking of a $Z_p$ symmetry. There are single phases with periodicity $p=2$, $p=3$ and $p=4$ with magnetization $m = 0, - \frac{1}{6}, -\frac{1}{4}$, respectively, and two phases with $p=5$ and magnetizations $m = -\frac{3}{10}$ and $m = -\frac{1}{10}$. The phases are arranged such that the magnetization as a function of $\Delta$ monotonically increases in a stepwise fashion. Each classical transition possesses a macroscopic degeneracy given by Eq.~\eqref{degeneracy} and thus a finite entropy that will be quenched by a finite hopping $J_\perp$. Furthermore, for a finite $J_\perp$ as shown in Fig.~\ref{fig:PDJ5-1} each classical transition serves as the origin of a fan containing a Luttinger liquid phase with an extension given by Eq.~\eqref{LLExt}. The edges of each fan are Lifshitz transitions where the Luttinger liquid parameter assumes the value given in Eq.~\eqref{KLpp}.

It is clear that for sufficiently large $J_\perp$ Luttinger liquid phases attributed to adjacent classical transitions will start to overlap (not shown). This has important consequence for the full Hamiltonian \eqref{Model} with the long-range interaction $J_{|r|}$.
As the extension of a commensurate state with periodicity $p$ on the $\Omega=0$ axis is on the order of $\delta \Delta \sim J_{p-1} + \mathcal{O}(J_{p})$ for $J_\perp = 0$,\cite{Bak82} it follows from Eq.~\eqref{LLExt} that for $J_{p-1} \lesssim J_\perp$ all higher commensurate phases of period $p$ will be washed out. This means that if, e.g., $J_{5} \lesssim J_\perp$, one crosses at $\Omega = 0$ as a function of $\Delta$ only 10 Lifshitz transitions as in Fig.~\ref{fig:PDJ5-1} even if the full long-range interaction $J_{|r|}$ of Eq.~\eqref{LongRangeInt} is considered. This simplifies the topology of the phase diagram in the $(\Delta,\Omega)$ plane of the full Hamiltonian \eqref{Model}  considerably.

\subsubsection{Phase diagram}
\label{sec:CPPhaseDiagram}

\begin{figure}
\includegraphics[width=0.4\textwidth]{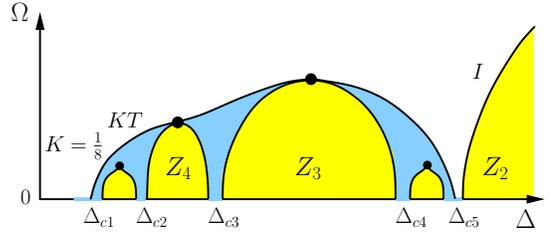}
\caption{Topology of the phase diagram for the Hamiltonian \eqref{Model} for a hopping $J_\perp \gtrsim J_{5}$ with commensurate phases breaking $Z_2$, $Z_3$, $Z_4$ and $Z_5$ (two smallest lobes) symmetries. Commensurate phases with higher periodicities are washed out. The positions of the critical fields $\Delta_{ci}$ are approximately given by Eqs.~\eqref{J5CrDelta}. The figure is not to scale; the extension of the commensurate phases with larger unit cells is exaggerated to make them visible.
 }
\label{fig:PDJ5-2}
\end{figure}

The phase diagram of the Hamiltonian \eqref{Model} with the long-range interaction $J_{|r|}$ in the $(\Delta,\Omega)$ plane defined by the Rabi frequency $\Omega$ and detuning $\Delta$ is shown in Fig.~\ref{fig:PDJ5-2}. For simplicity and illustrative purposes, we assume a finite hopping of Rydberg excitations on the order of $J_\perp \sim J_{5}$ so that commensurate phases with periodicity $p>5$ are not realized. Note that in typical optical lattice experiments the super-exchange interaction is usually
much smaller $J_\perp \sim J_p$, with $p$ reaching up to $10^3$ for deep optical lattices; see discussion in section \ref{sec:Exp}.
The restriction to $J_\perp \sim J_{p}$ with $p=5$ streamlines the discussion and captures all essential features with obvious generalizations to larger $p$. Apart from a broken $Z_2$ Ising phase, there are broken $Z_3$, $Z_4$ and $Z_5$ phases for such parameters whose corresponding spin configuration was illustrated already in Fig.~\ref{fig:PDJ5-1}. Up to corrections of order $J_5$, the critical fields $\Delta_{ci}$ are given by Eq.~\eqref{J5CrDelta}.

The phases on the axis $\Omega=0$ follow from a vertical cut through a phase diagram like that of the lower panel of Fig.~\ref{fig:PDJ5-1} at a finite $J_\perp \gtrsim J_{5}$, so that only 10 Lifshitz transitions are crossed giving rise to five Luttinger liquid phases (blue shaded) on the $\Omega = 0$ axis. Repeating similar arguments as in section \ref{sec:J1J2chain} about the stability of these Luttinger liquid phases with respect to the $S^x_i$ operator associated with a finite $\Omega$, we find that the broken $Z_2$ phase is bounded by an Ising transition line denoted by $I$. The other broken $Z_p$ phases are surrounded by a floating Luttinger liquid phase that extends to finite $\Omega$ and is itself bounded by a Kosterlitz-Thouless transition (KT). At the latter transition, the Luttinger liquid parameter assumes the universal value $K=1/8$. We will discuss the physics and the signatures of this KT transition in the next section in more detail.

The topology of the phase diagram for larger $\Omega$ follows from known results.\cite{Jose77,Coppersmith81,Ostlund81,Haldane83,Schulz83} First, the KT transition line close to $\Delta_{c1}$ connects with the one close to the Ising transition so that the Luttinger liquid phase houses all gapped commensurate phase of periodicity $p > 2$. Each broken $Z_p$ phase possesses a special multicritical transition point where the magnetization inside and outside the symmetry broken phase is accidentally commensurate. Whereas for the $Z_3$ and $Z_4$ phases this point connects directly with the polarized phase for large $\Omega$, it is located within the Luttinger liquid phase for the $Z_5$ phase (and also for higher commensurate phases that appear for smaller $J_\perp$). For periodicity $p=3,4$ this multicritical point is of Potts type,\cite{Jose77,Truong81,Schulz83} and for
the higher commensurate phases it is in the KT universality class with $K=2/p^2$, i.e.~the Luttinger parameter at the multicritical point is larger by a factor of two compared to the corresponding Lifshitz transition; see Eq.~\eqref{KLpp}.
The transitions from the commensurate phases into the Luttinger liquid away from these special points are instead Lifshitz transitions.\cite{Japaridze78,Pokrovsky79} The width $\delta \Delta$ and height $\delta \Omega$ of the dome housing the Luttinger liquid phases can be estimated to be on the order of the next-nearest neighbor coupling, $\delta \Delta \sim \delta \Omega \sim J_2 + \mathcal{O}(J_3)$.

\section{Dislocation-unbinding transition of the Rydberg crystal}
\label{sec:KT}

We now concentrate on the parameter regime of the phase diagram where both Rabi frequency, $\Omega$, and detuning, $\Delta$, are small compared to the hopping of Rydberg excitations, $J_\perp \gtrsim \Omega, |\Delta|$. We have seen in the last section that for increasing $\Delta$ a first commensurate state is stabilized for $\Omega=0$ only at $\Delta \sim J_\perp$; see Eq.~\eqref{LLExt}. We therefore expect that in the regime $\Delta \ll J_\perp$ a continuum description  that neglects any commensurability effects should be possible. Such a description is at the focus of this section. This allows us, in particular, to elucidate the nature of the floating Luttinger liquid phase and, in particular, the Kosterlitz-Thouless transition with Luttinger parameter $K=1/8$.

We start with representing the spin operators with the standard Jordan-Wigner fermions, $\{f^\dagger_i,f_j\} = \delta_{ij}$,
\begin{align}
S^+_i &= f^\dagger_i e^{- i \phi_i},\quad
S^-_i = e^{i \phi_i} f^\pdag_i,\\
S^z_i &= \hat{n}_i - \frac{1}{2} = f^\dagger_i f^\pdag_i - \frac{1}{2},\quad
\phi_i = - \pi \sum_{j<i} \hat{n}_j,
\end{align}
so that the Hamiltonian \eqref{Model} becomes
\begin{align}
\label{Model1}
\mathcal H =& \sum_{j <\ell} J_{|j-\ell|} \, \hat n_j \hat n_\ell
- \frac{J_\perp}{2} \sum_j \left(f^\dagger_j  f^\pdag_{j+1} + {\rm h.c.} \right)
\\\nn&
-\Delta \sum_j \left(\hat{n}_j - \frac{1}{2}\right) + \frac{\Omega}{2} \sum_j \left(f^\dagger_j e^{- i \phi_j}+e^{i \phi_j} f_j\right).
\end{align}

We will again apply the strategy that we analyze first the Hamiltonian for $\Omega = 0$, and then consider the effect of a finite Rabi frequency $\Omega$ perturbatively. For $\Omega=0$, we can distinguish between two regimes: whereas for $\Delta < \Delta^*$, with $\Delta^*$ to be determined below, the Rydberg excitations behave like a dilute weakly interacting gas, they form an incommensurate Rydberg crystal for $\Delta^* < \Delta \ll J_\perp$.

\subsection{Dilute weakly-interacting Rydberg gas}

Consider the Hamiltonian \eqref{Model1} for $\Omega=0$ at very small densities of Rydberg excitations $\langle \hat{n}_i \rangle \ll 1$. In this limit, there are only few Rydberg excitations that hardly interact. Neglecting this residual interaction, the dilute gas of Rydberg excitations can be approximated by the single-particle Hamiltonian
\begin{align} \label{EffGas}
\mathcal H_{\rm sp} =&
 \int dx\, %\left\{
 f^\dagger(x) \left( \frac{-\partial_x^2}{2m} - \mu\right) f(x),
\end{align}
where we have taken the continuum limit of small optical lattice constant, $a \to 0$, with $f_i \to f(x)/\sqrt{a}$, and $\hat n = f^\dagger f$. The mass $m$ and the chemical potential $\mu$ simply follow from the single-particle properties of Eq.~\eqref{Model1},
\begin{align} \label{EffParam}
\frac{1}{m} = J_\perp a^2,\quad
\mu = \Delta + J_\perp.
\end{align}
As the detuning $\Delta$ increases from negative values, the excitation gap, $-\mu$, for a single Rydberg excitation decreases and vanishes at
\begin{align} \label{Lifshitz}
\Delta^-_c = -J_\perp.
\end{align}
At $\Delta = \Delta_c^-$, there is a Lifshitz transition, the Rydberg excitations start to condense and form a Fermi sea. The critical value \eqref{Lifshitz} is in agreement with the results obtained in the last section; see Eq.~\eqref{LLExt}.
The energy density as a function of the density of Rydberg excitations $n_R$ is given by
the standard free Fermi gas expression
\begin{align} \label{E(n)1}
\varepsilon(n_R) = \frac{\pi^2 n_R^3}{6 m} - \mu n_R.
\end{align}
For $\mu > 0$, minimization with respect to the density gives $n_R = \sqrt{2m \mu}/\pi$ and, thus, the ground state energy of the Rydberg gas
\begin{align} \label{ERG}
\varepsilon_{RG} = - \frac{2 \sqrt{2}}{3\pi} \sqrt{m}\, \mu^{3/2}.
\end{align}
The residual interaction among Rydberg excitations neglected in Eq.~\eqref{EffGas} can in principle be extracted from the two-particle sector of the lattice Hamiltonian \eqref{Model1}. In the low-energy limit, we expect it to have the form $f^\dagger(x)(\partial_x f^\dagger(x))f(x)(\partial_x f(x))$, where the two gradients are required by the Pauli principle. This gives rise to a correction to Eq.~\eqref{E(n)1} on the order of $\mathcal{O}(n_R^4)$ that is sub-leading in the low-density limit $n_R \to 0$.

For small but finite densities, $\mu>0$, i.e., $\Delta > \Delta^-_c$, the small energy excitations on top of the Fermi sea are governed by the Luttinger liquid theory [see Eq.~\eqref{LuttModel}], characterized by the velocity $v$ and the Luttinger parameter $K$. They obey here the relation $v=v_F/K$ (which follows form the emergent Galilean symmetry), where $v_F = \sqrt{2\mu/m}$ is the single-particle velocity, and the Luttinger parameter derives from the compressibility
$\partial^2\varepsilon(n_R)/\partial n_R^2 = v \pi/K$ yielding
\begin{align} \label{RGasLLParameters}
K = 1 - \mathcal{O}(n_R).
\end{align}
The Luttinger parameter attains the value $K=1$ close to the Lifshitz transition, $n_R \to 0$, i.e., $\mu \to 0^+$ in agreement with Eq.~\eqref{KHc2} or, more generally, with Eq.~\eqref{KLpp}.

A Luttinger liquid with such a large parameter $K$ becomes immediately unstable in the presence of a $S^x$ perturbation, that is associated with the transverse field $\Omega$. The operator $S^x$ is relevant for $K>1/8$ [see Eq.~\eqref{KLHc1}], and thus opens a gap. So we obtain the result that in the small density limit, the gas of Rydberg excitations is only gapless for $\Omega = 0$ and gapped for any finite $\Omega$. We now turn to the situation of larger densities.

\subsection{Incommensurate Rydberg crystal}

If at $\Omega=0$ the detuning $\Delta$ is further increased, the density of Rydberg excitations increases and the interaction becomes more and more important. If $\Delta$ is sufficiently large, $\Delta \gtrsim \Delta^*$, the interaction $J_{|r|}$ finally cannot be treated perturbatively anymore. In order to determine the crossover value $\Delta^*$, we start now from the opposite limit and treat perturbatively the kinetic energy of Rydberg excitations, i.e., the coefficient $1/m$ in Eq.~\eqref{EffGas}.

\subsubsection{Rydberg crystal for $1/m = \Omega = 0$}

Consider the classical limit of the Hamiltonian \eqref{Model1}. In the continuum limit of small optical lattice constant, $a \to 0$, it reads
\begin{align}
\mathcal{H}_{\rm class} =
\frac{1}{2}  \int dx dy\,
\hat{n}(x) J_{|x-y|/a}  \hat{n}(y) - \mu \int dx\, \hat{n}(x),
\end{align}
with $\mu$ given in Eqs.~\eqref{EffParam}.
For $\mu > 0$ there is a finite number of particles, i.e., Rydberg excitations in the ground-state that interact via the long-range interaction $J_{|r|}$. In the thermodynamic limit $L \to \infty$, an eigenstate with a fixed density $n_R$ of Rydberg excitations is
\begin{align}\label{WCstate}
|\Psi_{\rm RC}\rangle = \prod^{\infty}_{j=-\infty}  S^+(x_j) |0\rangle,
\end{align}
where we abbreviated $S^+(x) = f^\dagger(x) e^{- i \phi(x)}$. The state $|\Psi_{\rm RC}\rangle$ depends on the positions, $x_j$,  that we assume to be ordered such that  $... < x_{j-1} < x_j < x_{j+1} < ...$.

The state $|\Psi_{\rm RC}\rangle$ is an eigenstate of the density operator $\hat{n}(x)$ with an eigenvalue given by the Dirac comb
\begin{align}
\hat{n}(x) |\Psi_{\rm RC}\rangle
= \sum_{j=-\infty}^{\infty} \delta\left(x - x_j \right)|\Psi_{\rm RC}\rangle.
\end{align}

\subsubsection{Ground state energy of the Rydberg crystal}

In the ground state $|\Psi^{(0)}_{\rm RC}\rangle$ the Rydberg excitations are equally spaced so that the positions $x_j$ of the wavefunction \eqref{WCstate} assume the values $x^{(0)}_j = j/n_R$,
\begin{align}\label{gWCstate}
|\Psi^{(0)}_{\rm RC}\rangle = \prod^{\infty}_{j=-\infty}  S^+(x^{(0)}_j)
|0\rangle.
\end{align}
The corresponding energy density $\varepsilon(n_R) |\Psi^{(0)}_{\rm RC}\rangle = \frac{1}{L} \mathcal{H}_{\rm class}|\Psi^{(0)}_{\rm RC}\rangle$ is a function of $n_R$ and reads
\begin{align}
\varepsilon(n_R) = \left(\sum_{m = 1}^{\infty} J_{|m|/(n_R a)} - \mu \right) n_R.
\end{align}
Using Eq.~\eqref{LongRangeInt} for the interaction, the energy simplifies to
\begin{align} \label{GroundstateFunction}
\varepsilon(n_R) &= J_R \zeta(\alpha) (n_R a)^\alpha n_R  - \mu n_R,
\end{align}
that has to be minimized in order to determine $n_R$, $\partial \varepsilon/\partial n_R = 0$,
\begin{align} \label{nRCrystal}
n_R = \frac{1}{a} \left(\frac{\mu}{J_R (\alpha+1) \zeta(\alpha)} \right)^{1/\alpha},
\end{align}
with $\zeta$ the Riemann zeta function. Taking $\varepsilon(n_R)$ at its minimum one finally obtains the ground-state energy for the Rydberg crystal state,
\begin{align} \label{GroundstateEnergy}
\varepsilon_{RC} = -\frac{\alpha}{(\alpha + 1) [(1+\alpha) \zeta(\alpha)]^{1/\alpha}}  \frac{\mu^{1+1/\alpha}}{J_R^{1/\alpha} a},
\end{align}
that is non-perturbative in the interaction $J_R$.

Comparing the energies of the Rydberg gas, Eq.~\eqref{ERG}, and the Rydberg crystal, we obtain for the crossover value of the detuning
\begin{align} \label{Xover}
\Delta^* = \mu^* + \Delta_c^-,\quad {\rm with}\quad
%\mu^* \sim \left[\frac{J_\perp^{\alpha/2}}{J_R} \right]^{\frac{2}{\alpha-2}}
\mu^* \sim J_\perp \left[\frac{J_\perp}{J_R} \right]^{\frac{2}{\alpha-2}},
\end{align}
and $\Delta_c^- = -J_\perp$.
Note that $\Delta^* < 0$ for $J_\perp \ll J_R$ and, consequently, the incommensurate Rydberg crystal exists in a finite regime $\Delta^* < \Delta < J_\perp$ between the Rydberg gas at smaller and the commensurate Rydberg crystal at larger detuning $\Delta$.

\subsubsection{Degenerate perturbation theory in the kinetic term}

The incommensurate Rydberg-crystal ground state of Eq.~\eqref{gWCstate} breaks the translational symmetry of the continuum Hamiltonian.
Consequently, quantum fluctuations induced by a finite hopping $J_\perp$ will give rise to phonon excitations.

Besides the simple shift in the expression for the chemical potential, $\mu$, a finite $J_\perp$ results in a kinetic energy with mass $m$ as defined in Eq.~\eqref{EffParam}. Consider the crystalline wavefunction in Eq.~\eqref{WCstate} where the positions of the excitations $x_j$ differ from their equilibrium positions by $u_j = x_j - x_j^{(0)}$. The eigenvalue problem in the presence of the kinetic term, $[\mathcal{H}_{\rm class} + \int dx f^\dagger (-\partial_x^2/(2m)) f - E] |\Psi_{\rm RC}\rangle = 0$, then reduces to finding the solution of the {\it first-quantized} Hamiltonian
\begin{align} \label{1stquantHam}
H =& \frac{1}{2}  \sum_{i \neq j} \,J_{|x_i-y_j|/a} - \mu N_R + \sum_j \frac{p^2_j}{2 m},
\end{align}
where the momentum and position obey the canonical commutation relation $[x_j, p_i] = i \delta_{ij}$, and $N_R = n_R L$.

Expanding the Hamiltonian in second order in the small deviations $u_j$ we obtain $H \approx L \varepsilon(n_R) + H_{\rm phonon}$ with the energy $\varepsilon(n_R)$ of Eq.~\eqref{GroundstateFunction} and the phonon Hamiltonian
\begin{align} \label{Hphonons}
H_{\rm phonon} = \sum_j
\frac{p_j^2}{2m}+ \frac{1}{4} \sum_{i\neq j} J''_{|i-j|/(n_R a)}  \frac{1}{a^2}  (u_i - u_{j})^2,
\end{align}
where $J''_{|r|} = \partial^2J_{|r|}/\partial |r|^2$.
In order to obtain the asymptotics at lowest energies, we consider a second continuum limit, i.e., we confine  ourselves to length scales much larger than the mean distance between Rydberg excitations, $1/n_R$. In this limit, we can use $u_i \to u(x)$ and $p_i \to p(x)$, with the commutation relation $[u(x),p(x')] \approx i \delta(x-x')/n_R$, and the phonon Hamiltonian is approximated by
\begin{align} \label{Hphonons2}
H_{\rm phonon} \approx n_R \int dx  \left[
\frac{p^2}{2m}+ \frac{1}{2} \frac{\mathcal{B}}{(a n_R)^2} (\partial_x u)^2
\right],
\end{align}
where we kept only the lowest derivatives of the $u(x)$ field and we abbreviated
\begin{align}
\mathcal{B} = \sum_{m=1}^\infty J''_{|m|/(n_R a)} m^2 =
\sum_{m=1}^\infty \frac{ \alpha( \alpha+1) J_R m^2 }{(|m|/(n_R a))^{\alpha+2}}
\\\nn
= \alpha( \alpha+1) J_R (n_R a)^{2+\alpha} \zeta(\alpha).
\end{align}
Introducing the dimensionless fields $u(x) = - \phi(x)/(\pi n_R)$ and $p(x) = - \partial_{x}\theta(x)$ with the canonical commutation relations $[\frac{1}{\pi}  \phi(x), \partial_{x'}\theta(x')] = i \delta(x-x')$, the phonon Hamiltonian \eqref{Hphonons2} can be recast in the standard Luttinger liquid form of Eq.~\eqref{LuttModel} with velocity $v$ and Luttinger parameter $K$ obeying
\begin{align} \label{KV}
K v &= \frac{\pi n_R}{m},\quad v^2 = \frac{1}{m} \frac{\mathcal{B}}{(a n_R)^2}.
\end{align}
Using the explicit expressions for $\mathcal{B}$, for the mass $m$ given in Eq.~\eqref{EffParam}, and Eq.~\eqref{nRCrystal} for $n_R$ we obtain
\begin{align}
v &\sim \sqrt{J_\perp a^2 \mu},
\\\label{LLParameters}
K &\sim \sqrt{\frac{J_\perp}{\mu}} \left(\frac{\mu}{J_R}\right)^{1/\alpha} \sim \left(\frac{\mu^*}{\mu}\right)^{\frac{1}{2}-\frac{1}{\alpha}}.
\end{align}
The Luttinger parameter is small, $K \ll 1$, in the incommensurate Rydberg crystal regime $\mu^* \ll \mu $. It increases and becomes of order one if the chemical potential is lowered down to the crossover value $\mu^*$ of Eq.~\eqref{Xover}, and it thus smoothly connects with Eq.~\eqref{RGasLLParameters} of the dilute Rydberg gas regime.

As a consistency check, we might consider the correction to the ground state energy due to the phonon excitations. This is of order $\delta \varepsilon \sim v \int dk k$ where the momentum integral is cutoff at $n_R$ so that one obtains for the relative correction
\begin{align}
\frac{\delta \varepsilon}{\varepsilon_{RC}} \sim \sqrt{\frac{J_\perp}{\mu}} \left(\frac{\mu}{J_R} \right)^{1/\alpha}
\sim \left(\frac{\mu^*}{\mu}\right)^{\frac{1}{2}-\frac{1}{\alpha}}.
\end{align}
Note that the relative correction is on the order of the Luttinger parameter $\delta \varepsilon/\varepsilon_{RC} \sim K$.
So the correction is of order one and the perturbative treatment in the collective variables $u_i$ and $p_i$ breaks down if the chemical potential is lowered down to the crossover value $\mu^*$ of Eq.~\eqref{Xover}.

\subsubsection{Stability analysis with respect to $\Omega$ perturbations: dislocation-mediated melting transition }
\label{sec:KTMelting}

\begin{figure}
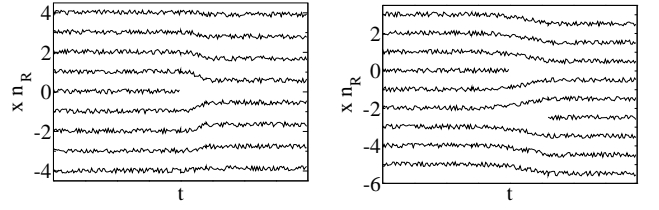

\includegraphics[width=0.22\textwidth]{fig5a} \quad
\includegraphics[width=0.22\textwidth]{fig5b}
\caption{Worldlines of the incommensurate Rydberg crystal. Vertical axis shows position $x$ in units of $1/n_R$ and horizontal axis shows time $t$. The wiggles indicate phonon excitations. The left panel shows a single dislocation and the right panel a dislocation pair. The unbinding of such pairs leads to a melting of the crystal at the KT transition.}
\label{fig:RC}
\end{figure}

In the presence of a finite Rabi frequency $\Omega$ Rydberg excitations can be added to or removed from the crystalline arrangement. In the limit $a\to 0$, with $f_i \to f(x)/\sqrt{a}$, this perturbation can be written as
\begin{align} \label{RabiPert}
\delta \mathcal H = \frac{\Omega}{2\sqrt{a}} \int dx \left(S^+(x) + S^-(x) \right),
\end{align}
where $S^+(x) = f^\dagger(x) e^{- i \phi(x)}$ and $S^-(x) = (S^+(x))^\dagger$. In the thermodynamic limit,
the operation of $S^+(x)$ on the Rydberg-crystal wavefunction, Eq.~\eqref{WCstate}, i.e., adding a Rydberg excitation can be expressed in terms of a {\it first-quantized} operator (neglecting Klein factors),
\begin{align}
S^+(x) \sim \frac{1}{\sqrt{2\pi a}} : e^{i (x - x_n) p_n + \sum^{n-1}_{j=-\infty} i (x_{j+1} - x_j) p_j}:,
\end{align}
where the normal ordering, $:\hat O:$, is defined such that the momentum operators should be placed to the right of all position operators. The index $n$ for a given $x$ is defined such that $x_n < x < x_{n+1}$.
As we require that the particle positions are ordered, the addition of an excitation at position $x$ involves a collective shift of all particles with positions $x_j < x$, that is just realized by the sum of momentum operators in the exponent. The above expression simplifies in the continuum limit,
\begin{align}
S^+(x) \sim \frac{1}{\sqrt{2\pi a}} e^{i \int^x_{-\infty} dy p(y)} = \frac{1}{\sqrt{2 \pi a}} e^{- i \theta(x)},
\end{align}
where we used the continuum version of the momentum operator and its relation to the $\theta$ field as defined in the last paragraph. We have also set $\theta(-\infty) = 0$.
The perturbation \eqref{RabiPert} on the Rydberg crystal thus corresponds to the first-quantized operator
\begin{align} \label{DisCreator}
\delta H = \frac{\Omega}{\sqrt{2\pi} a} \int dx \cos (\theta ),
\end{align}
[see also Eq.~\eqref{KLHc1}]. This operator can create or destroy dislocations in the incommensurate Rydberg crystal.

Fig.~\ref{fig:RC} shows the worldlines of Rydberg excitations. In the ground state these worldlines are on average equally spaced with a distance given by $1/n_R$. The wiggles around the equilibrium positions illustrate the small-amplitude, phonon excitations. If the operator $S^-$ is applied on the Rydberg crystal, a Rydberg excitation is destroyed at a certain time and a dislocation is created; see left panel of Fig.~\ref{fig:RC}. The remaining excitations will accommodate themselves at a later time but the averaged distance has slightly increased and, as a consequence, the resulting state is energetically unfavorable. In order to recover the ground-state configuration, a particle has to be added at a later time by applying the $S^+$ operator giving rise to an anti-dislocation as shown in the right panel. If the incommensurate Rydberg crystal is stable, the dislocation and anti-dislocation in the worldlines are bound and, in contrast to the phonons, correspond to large-amplitude excitations. This is the case if the operator of Eq.~\eqref{DisCreator} is irrelevant in the RG sense which implies $K \leq 1/8$.\cite{GiamarchiBook} As the Luttinger parameter $K \ll 1$ for $\mu \gg \mu^*$, the incommensurate Rydberg crystal is stable to such small $\Omega$ perturbations. The dislocation pairs, however, unbound for $K>1/8$ and, as a consequence, the Rydberg crystal melts in a Kosterlitz-Thouless transition. We can estimate the position of the KT transition using Eq.~\eqref{LLParameters}. The Luttinger parameter becomes of order one for $\mu \sim \mu^*$, and the location of the transition thus coincides with the crossover between the Rydberg gas and the incommensurate Rydberg crystal. As we argued previously, from the separatrix of the RG flow for the KT transition\cite{Kosterlitz73} follows that the phase boundary starts linearly in $\Omega$, i.e., $\Delta_{\rm KT}(\Omega)-\Delta_{\rm KT}(0) \sim \Omega$.

\section{Summary  and Discussion}
\label{sec:summary}
Combining the results of section \ref{sec:CommPhases}  and \ref{sec:KT}, we summarize in the following the topology of the phase diagram
for one-dimensional Rydberg atoms,
that are governed by the Hamiltonian \eqref{Model}. We first outline the global structure of the phase diagram before we focus on the experimentally relevant regime and discuss possible experimental consequences.

\subsection{Phase diagram}

\begin{figure}
\includegraphics[width=0.3\textwidth]{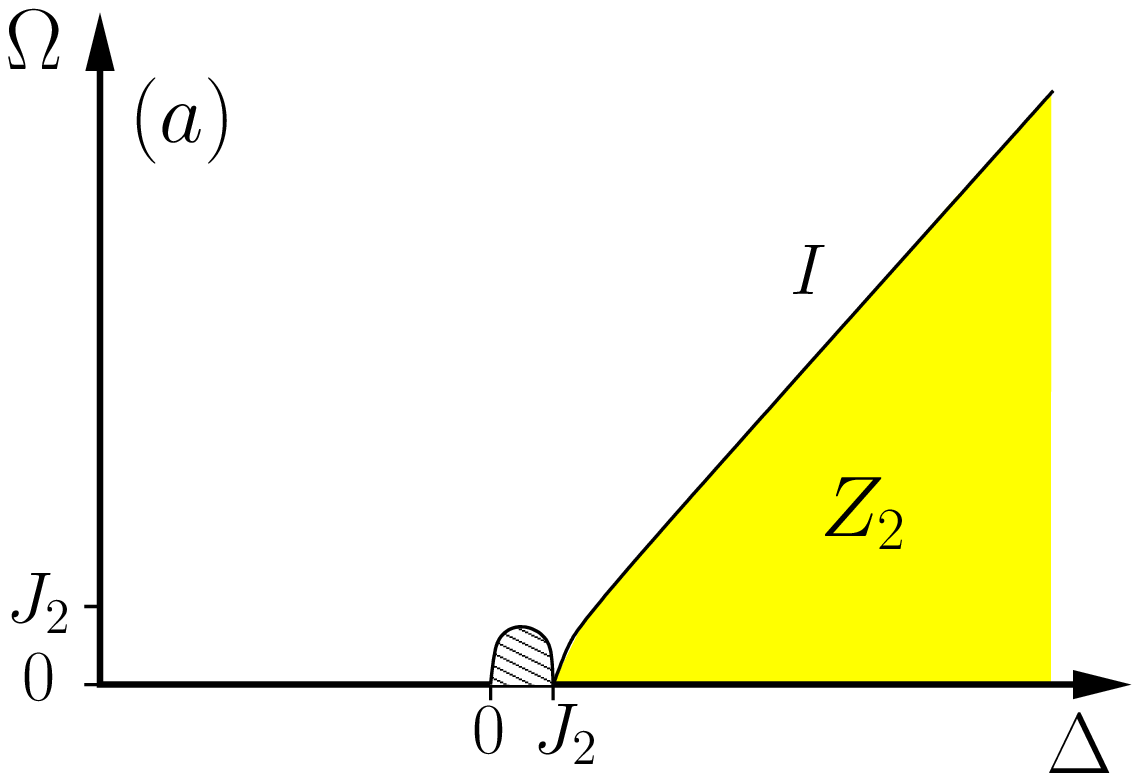}
\includegraphics[width=0.32\textwidth]{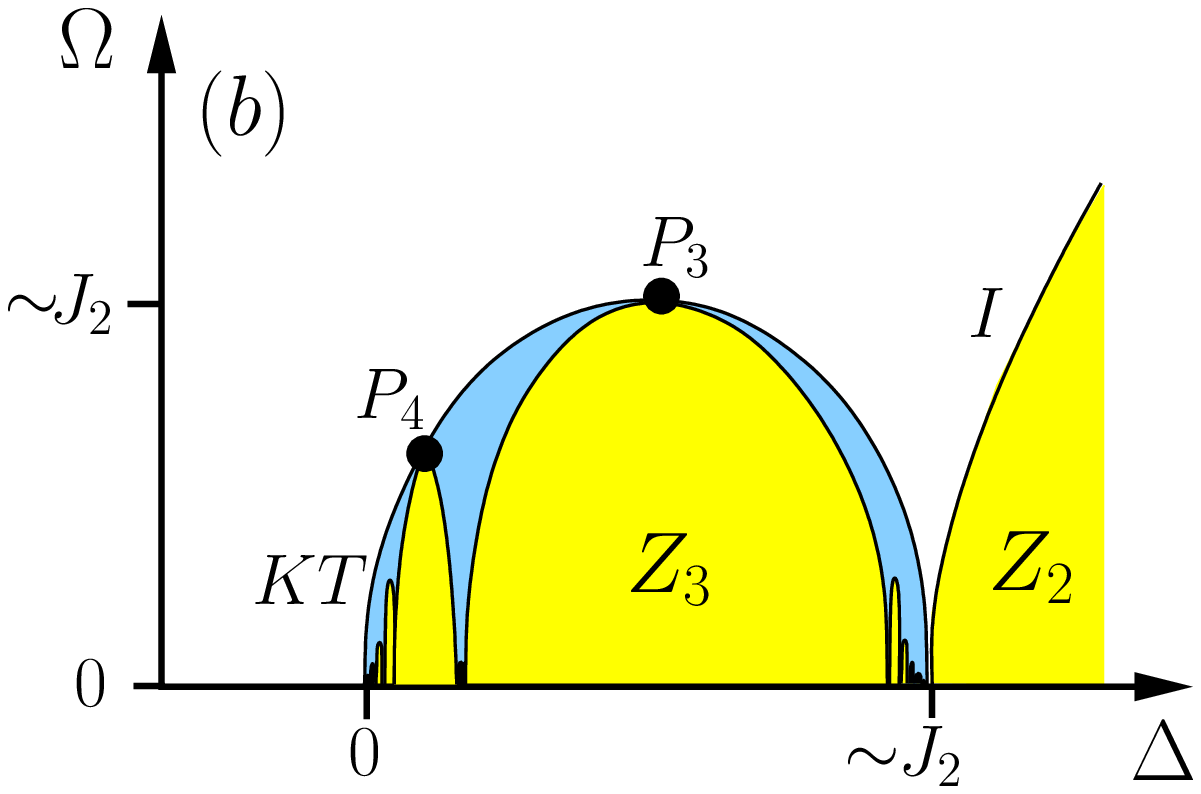}
\includegraphics[width=0.3\textwidth]{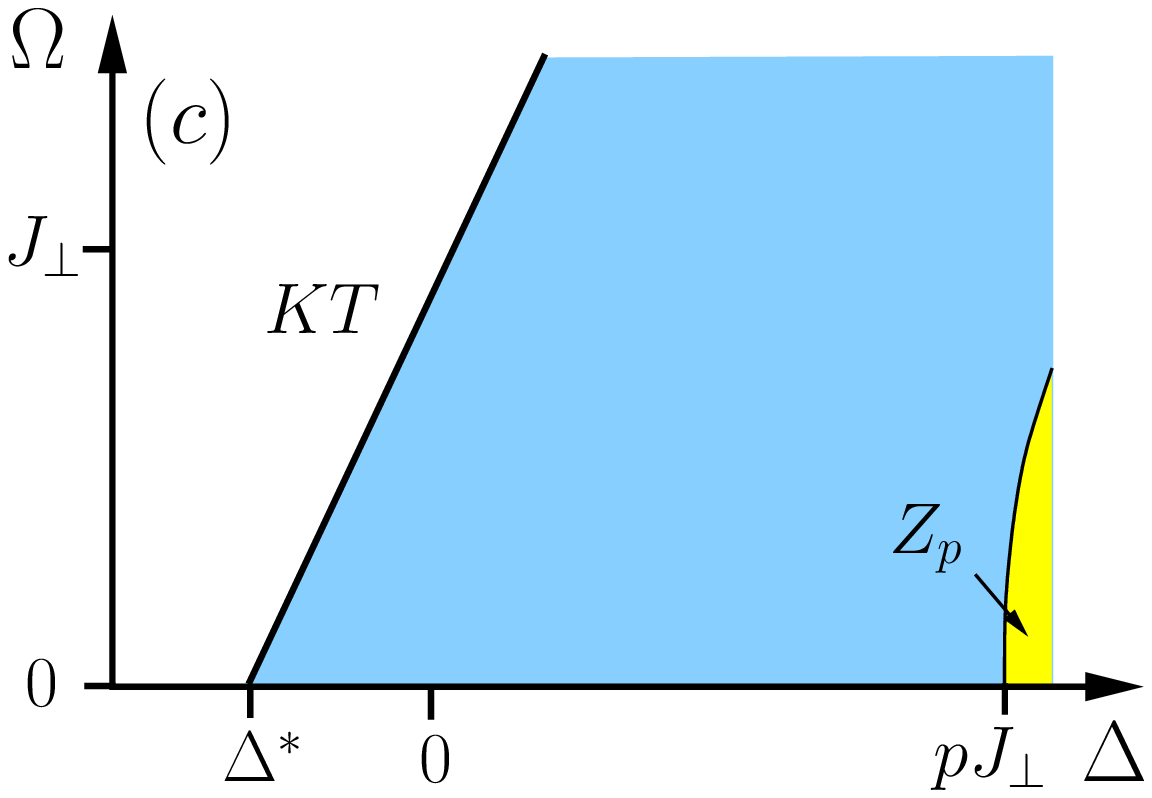}
\caption{Sketch of the phase diagram in the $(\Delta,\Omega)$ plane of the Hamiltonian \eqref{Model} close to the origin but on various scales: (a) for $|\Delta|, \Omega \gg J_2$, (b)
for $J_\perp \ll |\Delta|, \Omega \lesssim J_2$
and (c) $|\Delta|, \Omega \lesssim J_\perp$; see text. There are gapped commensurate (yellow), gapless incommensurate (blue) phases, and a gapped polarized phase (white). The commensurate phases break a $Z_p$ symmetry of the optical lattice and their extension decreases fast with the order of commensurability $p$. The commensurability of the last commensurate phase at small $\Delta$ in (c)  is determined by $J_p \sim J_\perp$, i.e. $p \sim (J_R/J_\perp)^{1/\alpha}$. The abbreviations $I$ and $KT$ indicate an Ising transition and a Kosterlitz-Thouless transition line, respectively. Points $P_n$ denote critical points of $n$-state Potts universality.
}
\label{fig:PhaseDiagram}
\end{figure}

The topology of the phase diagram depends qualitatively on the relation of $\Delta$ and $\Omega$ with respect to the hopping $J_\perp$ and, especially, the next-nearest neighbor interaction $J_2 = J_R/2^\alpha$; see Fig.~\ref{fig:PhaseDiagram}.

On scales $J_\perp, J_2 \ll |\Delta|, \Omega \ll J_R$ [see Fig.~\ref{fig:PhaseDiagram}(a)], the effect of hopping $J_\perp$ and higher order interactions $J_{|r|}$ with $r>1$ is negligible and the Hamiltonian \eqref{Model} reduces effectively to an Ising model in a longitudinal and transverse magnetic field. Consequently, one obtains a single Ising transition located\cite{Ovchinnikov03} at $\Omega \propto \Delta$ between a paramagnet and an antiferromagnet with a Rydberg excitation at every second lattice site, which breaks a $Z_2$ symmetry of the lattice. For smaller $\Omega \lesssim J_2$ the Ising transition line, however, deviates from the straight line and approaches a value on the $\Omega = 0$ axis on the order $\Delta \sim J_2$ [neglecting corrections of order $\mathcal{O}(J_3)$].

At intermediate scales $J_\perp \ll |\Delta|, \Omega \sim J_2$, an additional dome close to the Ising transition appears that contains higher commensurate phases. Fig.~\ref{fig:PhaseDiagram}(b) shows clearly the presence of a lobe with a broken $Z_3$ phase where a Rydberg excitation is present at every third site of the optical lattice. The width and height of the lobes decreases fast with the order of commensurability so that higher-order commensurate phases are barely visible on the scale of Fig.~\ref{fig:PhaseDiagram}(b). In fact, most of the commensurate phases of high order are not realized due to the finite hopping $J_\perp$. As a rule of thumb, one can say that only those commensurate phases breaking a $Z_p$ symmetry of the optical lattice appear in the $(\Delta, \Omega)$ plane
which fullfill $J_\perp < J_{p-1}$; all phases with higher commensurabilities $p'$ with $J_{p'-1} \ll J_\perp$ are washed out; see discussion in section \ref{sec:pp'Transitions}.  In addition, the extension of the broken $Z_p$ phases on the $\Omega=0$ axis is approximately on the order of $\delta \Delta \sim J_{p-1} + \mathcal{O}(J_{p})$, and thus strongly decreases with increasing $p$. The regions between these commensurate phases is filled with an incommensurate floating phase. Here, the Rydberg crystal has a periodicity that is incommensurate with the underlying optical lattice and, in addition, supports low-energy phonon excitations described by the Luttinger liquid theory. The existence of this floating phase was already pointed out in Ref.~\onlinecite{Weimer}.
This floating phase is microscopically stabilized by the hopping of Rydberg excitations. Although the bare value of the hopping is very small, $J_\perp \ll |\Delta|, \Omega$, it is strongly enhanced by renormalization effects generated by the laser drive $\Omega$,\cite{Weimer} so that the floating phase even extends to large values of $\Omega \lesssim J_2$. This in fact implies that the phase diagrams in Figs.~\ref{fig:PhaseDiagram}(a) and (b) even apply to the model Eq.~(\ref{Model}) in the limit $J_\perp \to 0$,
\begin{align}
\label{nojperp}
\mathcal H_{J_\perp \to 0}=& \sum_{\ell > j} J_{|\ell-j|} \, \big(S_j^z+1/2\big) \big(S_\ell^z+1/2 \big)
\\\nn&
-\Delta \sum_j S_j^z + \Omega \sum_j S_j^x.
\end{align}
In this limit the full fractal structure of the classical long range Ising model is restored.

The transition between the incommensurate to a certain commensurate Rydberg crystal state is in the usual universality class of commensurate-incommensurate (C-IC) transitions\cite{Japaridze78,Pokrovsky79} except at special points of accidental commensurabilities. Furthermore, we find that the boundary of the dome is given by a Kosterlitz-Thouless (KT) transition where the incommensurate Rydberg crystal melts due to the proliferation of dislocations.\cite{Nelson,Young} The Luttinger parameter at this KT transition takes the universal value $K=1/8$ with the concomitant universal correlation tails for Rydberg excitations. There are two special multicritical point on the boundary of the dome where the magnetization is accidentally commensurate with the $Z_3$ and $Z_4$ phases denoted by $P_3$ and $P_4$, respectively. At these two points, the transition is not of KT type but instead in the three- and four-state Potts universality class.\cite{Haldane83,Schulz83}

Finally, at very small scales $|\Delta|, \Omega \ll J_\perp$ [see Fig.~\ref{fig:PhaseDiagram}(c)], commensurability effects are negligible and an effective continuum description is possible; see section \ref{sec:KT}. This allows to describe the physics at the lower left edge of the dome where the KT transition hits the $\Omega=0$ axis. The floating incommensurate phase can be here captured within a standard strong-coupling perturbation theory familiar, for example, from the treatment of Wigner crystals. The phonons of the incommensurate Rydberg crystal give rise to gapless excitations and the Luttinger liquid parameter can be determined. We find that the creation of dislocations by the Rabi frequency $\Omega$ is irrelevant and the floating phase is stable as long as the detuning exceeds $\Delta > \Delta^*$, where $\Delta^*=-J_\perp + \mu^*$ with $\mu^* \sim J_\perp (J_\perp/J_R)^{2/(\alpha-2)}$. For smaller detuning, the Luttinger parameter reaches $K=1/8$ and a dislocation-mediated (KT) melting of the incommensurate Rydberg crystal sets in. If $\Delta$ increases to values of order $p J_\perp$ [see Eq.~\eqref{LLExt}], where $p\sim (J_R/J_\perp)^{1/\alpha}$ follows from the criterion $J_p \sim J_\perp$, the floating phase becomes unstable as well and a first commensurate state with periodicity $p$ takes over.

\subsection{Experimental parameters and signatures
of the two melting transitions}
\label{sec:Exp}

In order to estimate the location of a typical experiment on Rydberg atoms in the phase diagram we now discuss typical values for parameters. In case of a van der Waals interaction between Rydberg atoms the largest energy scale $J_R$ is given by $J_R = C_6/a^6$ with $C_6 \approx 500$GHz$\,\mu m^6$ for heavy alkali atoms with principal quantum number of order $n \approx 70$.\cite{Walker08} If the interaction is instead of dipolar type one has $J_R = V_{dd}/a^3$ where $V_{dd} \approx 1$GHz$\,\mu m^3$ for heavy alkalis with $n \approx 15$.\cite{Schachenmayer}
Using an optical lattice constant $a \approx 0.5 \mu m$ one obtains $J_R \approx 32$THz
% 1THz == 7.6 Kelvin
and $J_R \approx 8$GHz for the van der Waals and the dipolar interaction, respectively. Rabi frequencies of order $\Omega \approx 100$kHz have been achieved\cite{Low} that is on the order of the $r^{\rm th}$ nearest neighbor interaction, $J_r = J_R/r^\alpha$, with $r \approx 26$ and $r \approx 43$ for the two cases, respectively. On the other hand, for a deep optical lattice with potential depth of $20 E_r$ we can estimate the hopping to be $J_\perp \approx 10^{-6} E_r$.\cite{BDZ}   For this rough estimate we've assumed that the on-site interaction energy between Rydberg and ground-state atoms is of the same order as the interaction energy between two ground-state atoms.  For a typical recoil energy $E_r \approx 10$kHz this yields $J_\perp \approx 10^{-2}$Hz that is very small compared to all other energy scales, justifying the approximation of a frozen Rydberg gas.

Taken together, this would place a typical experiment in the parameter regime pictured in Fig.~\ref{fig:PhaseDiagram}(b). An important requirement for the observation of different phases in this phase diagram is, however, that the corresponding ground state energy is much larger than the decay rate of the Rydberg excitations, $1/\tau$. In particular, the life-time $\tau$ should be much larger than the time of several Rabi cycles in order to enable the crystallization of Rydberg excitations. Even then, it will be very difficult to identify experimentally the universality classes of transitions {\it between} different phases of Fig.~\ref{fig:PhaseDiagram} because due to critical slowing down very long life-times of Rydberg excitations would be required, especially, to distinguish between the Kosterlitz-Thouless and a possible first order transition.
Moreover, it is important to note that experiments
with quasi one-dimensional atomic gases usually operate in a configuration where a large number of independent, parallel 1D tubes are arranged in a 2D array with the number of atoms per tube typically around $\sim 100$. In order to observe a crystalline arrangement of Rydberg excitation, one needs to excite a sizable fraction of the atoms, say, about $1/10$, so that one has to operate in a regime where the parameters $\Omega \sim \Delta \sim J_R/10^\alpha$ are accessible.
For experimental protocols for cooling the system in order to realize Rydberg crystallization we refer the reader to the literature.\cite{Pohl,Schachenmayer}

The important finding of this work is that as function of increasing detuning $\Delta$ at fixed $\Omega$ one approaches first a Kosterlitz-Thouless transition to enter the regime of an incommensurate Rydberg crystal. In principle, this KT melting transition can be detected by Bragg scattering of light, which can be achieved by a resonant de-excitation of the Rydberg atoms.
If the Bragg signal  is measured along a 1D tube axis, it is just proportional to the 1D static structure factor
\begin{equation}
S(q) = \int dx \, e^{i q x} \langle S^z(x) S^z(0) \rangle,
\end{equation}
where $\lambda=2 \pi/q$ is the wavelength of the light and the spin, $S^z = n_R - \frac{1}{2}$, is related to the density of Rydberg excitations $n_R$. In the Rydberg gas phase, the structure factor does not exhibit any Bragg peaks at some finite momentum $q$. However, as the KT transition is reached and one enters the floating crystal phase (blue shaded in Fig.~\ref{fig:PhaseDiagram}), quasi-Bragg peaks will appear. From the algebraic decay of correlations \eqref{FCcorrel} in the floating incommensurate Rydberg crystal phase, one obtains
\begin{equation}
S(q) \sim \frac{1}{|q+ 2 \pi n_R|^{1-2K}} +\frac{1}{|q-2 \pi n_R|^{1-2K}} \ ,
\end{equation}
where the Luttinger parameter assumes its maximal value $K=1/8$ at the KT transition. The structure factor is thus expected to show a quasi-Bragg peak if the wavelength of the light matches the lattice spacing of the floating Rydberg crystal. As the quantum fluctuations are even further suppressed so that one reaches the C-IC transition and a commensurate Rydberg crystals forms (yellow shaded in Fig.~\ref{fig:PhaseDiagram}), true long-range crystalline order finally gives rise to well-defined (delta-function) Bragg peaks. In this way, the two melting transitions could be experimentally detected.

\acknowledgments

We thank W. Zwerger for suggesting this problem and for useful discussions. We also acknowledge interesting discussions with H.P. B\"uchler, I. Mekhov, who suggested the use of Bragg scattering to detect Rydberg crystals, and R. G. Pereira. E. S. is supported by the A. V. Humboldt foundation. M. P. is supported by the Erwin-Schr\"odinger-Fellowship J 3077-N16 of the Austrian Science Fund and acknowledges support by FOR801 of the DFG during the early stages of this work.
M.G. is supported by SFB 608 and FOR 960 of the DFG.

 %%%%%%%%%%%%%%%%%%%%%%%

\appendix

\section{Driven Rydberg atoms in the superfluid regime}
\label{app:BEC}

So far we have discussed the ground state phase diagram of driven Rydberg atoms in a one-dimensional optical lattice deep in the Mott insulating regime, where the Hamiltonian \eqref{hamHE} can be mapped to the effective spin model \eqref{Model}.
Virtual hopping processes, $t_\sigma$, via high-energy states involving doubly occupied sites contribute to the nearest-neighbor spin-exchange. Evaluation of this contribution in second order in the hopping $t_\sigma$ and in zeroth order in $\Omega$ one obtains
\begin{align} \label{Jperp}
J_\perp &= \frac{4 t_\uparrow t_\downarrow}{U_{\uparrow \downarrow}},
\\
J_z &= \frac{2 (t_\uparrow^2 + t_\downarrow^2)}{U_{\uparrow \downarrow}} - \frac{4 t_\uparrow^2}{U_{\uparrow \uparrow}} - \frac{4 t_\downarrow^2}{U_{\downarrow \downarrow}}.
\end{align}
In Eq.\ \eqref{Model}, we neglected the longitudinal part $J_z$ because it only gives rise to a negligible correction of the Rydberg-Rydberg interaction on nearest neighbor sites, which is several orders of magnitude larger. At this order one obtains the standard expression for the generated spin exchange.\cite{Kuklov, Duan} As the initial and the virtual states have the same number of up and down spin, the energy differences appearing in the denominators do not depend on the laser detuning $\Delta$.
The correction to these expressions at finite $\Omega$ originates from exciting and de-exciting a Rydberg atom or vice versa while in the virtual state it is located on a doubly occupied site. This yields a correction on the order of $\mathcal{O}\left(\frac{t^2 \Omega^2}{U^2 \Delta} \right)$. For $\Omega > |\Delta|$, this correction becomes large and will dominate over Eq.~\eqref{Jperp}. However, it turns out that in this parameter regime, the spin-exchange $J_\perp$ is strongly renormalized by processes {\it within} the low-energy Hilbert space so that its bare value can in fact be neglected. If two neighboring atoms are both excited or de-excited with $\Omega$ an effective $J_\perp$ on the order of $\Omega^2/\Delta$ is generated\cite{Weimer} that does not involve doubly occupied sites and, therefore, is much larger than the bare one. For our purposes, we can therefore restrict ourselves to the lowest order expression \eqref{Jperp}.

Now we briefly consider the opposite limit, where the ground-state atoms are in the superfluid regime and form a (quasi-)condensate which has been addressed in a recent experiment on Rydberg atoms in a 1D lattice.~\cite{Viteau}
In this case, we can apply the replacement $b_{j \downarrow} \to \sqrt{n_0+\sigma_{j\downarrow}} e^{-i \theta_{j \downarrow}}$ with the mean density of ground-state atoms, $n_0 \gg n_R$, assumed to be much larger than the number of excited Rydberg atoms $n_R$,
and $\sigma_{j \downarrow}$ and $\theta_{j \downarrow}$ account for fluctuations of the density and phase, respectively. In the 1D case, these fluctuations destroy true long-range order leaving only a quasi-condensate of ground-state atoms with algebraic correlations. The part of the Hamiltonian describing the the excited Rydberg atoms reads
\begin{align}
\label{h_sf}
\mathcal H =& -t_\uparrow \sum_{j} \big( b^\dagger_{j \uparrow} b^{\ }_{j+1 \uparrow} + \text{h.c.} \big)
+ \frac{1}{2}\sum_{j\neq\ell} J^{\ }_{|j-\ell|} \, b^\dagger_{j \uparrow} b^\dagger_{\ell \uparrow} b^{\ }_{\ell \uparrow} b^{\ }_{j \uparrow}
\nn\\
&+ \frac{\Omega }{2} \sum_j \big( b_{j \uparrow}^\dagger \sqrt{n_0+\sigma_{j\downarrow}} e^{-i \theta_{j \downarrow}}+ b_{j \uparrow}e^{i \theta_{j \downarrow}}\sqrt{n_0+\sigma_{j\downarrow}} \big)
\nn\\&
- \frac{1}{2} \sum_j \left(\Delta - 2 U_{\uparrow\downarrow} (n_0 + \sigma_{j,\downarrow}) \right)n_{j \uparrow}  ,
\end{align}
In writing the above Hamiltonian, we assumed a strong on-site repulsion $U_{\uparrow\uparrow}$ between excited Rydberg atoms which implies the hard-core constraint $n_{j\uparrow} \leq 1$. If we were to neglect the fluctuations of the (quasi)-condensate, $\theta_{j \downarrow}=\sigma_{j \downarrow}=0$, the above Hamiltonian could be mapped to the same pseudospin-Hamiltonian of Eq.\ \eqref{Model} with $J_\perp \to  2 t_{\uparrow}$ and $\Omega \to \sqrt{n_0} \, \Omega$ and $\Delta \to \Delta-2 U_{\uparrow\downarrow} n_0$. Within this approximation, we would expect the same ground-state phase diagram as the driven Mott-insulator. The main effect of the condensate is a scaling of the Rabi-frequency $\Omega$ with $\sqrt{n_0}$, as observed experimentally.\cite{Viteau} In general, however, the fluctuations will modify the phase diagram. For $\Omega=0$, we expect the density fluctuations $\sigma_i$ to influence the phase diagram only quantitatively and not qualitatively at least for incommensurate fillings $n_0$ because $\sigma_i$ couples to an oscillating Rydberg density profile $n_{j \uparrow}$.
At finite $\Omega$, on the other hand, the presence of phase fluctuations $\theta_{j \downarrow}$ reduce the relevancy of $\Omega$ operator in the renormalization group sense, so that the incommensurate Rydberg crystal will be stabilized. In addition, the Luttinger parameter at the Kosterlitz-Thouless melting transition, see Fig.~\ref{fig:PhaseDiagram}, will differ from $K=1/8$.
The influence of fluctuations on the Ising transition of Fig.~\ref{fig:PhaseDiagram}a is probably more complicated and left for a future study.

\section{General particle description for general  transitions between commensurate crystal phases}
\label{app}
In this appendix we provide a general definition of the fictitious particles introduced in Sec.\ref{sec:pp'Transitions}, and show that they form a subspace that becomes the degenerate ground state manifold at the classical transitions between commensurate states. We consider finite range interaction specified by Eq.~(\ref{frInt}).
Before diving into technicalities, we outline our main steps and results.  From the results of Hubbard,~\cite{Hubbard78} and Bak and Bruinsma,~\cite{Bak82} it follows that all states with fraction  $q/p$ of up spins with $p \le n+1$ [$n$ is the interaction range in Eq.~(\ref{frInt})] have a finite stability region and occur in increasing order upon increasing the longitudinal field $\Delta$. Given the fraction $q/p$, in order to minimize the interaction Eq.~(\ref{frInt}), the up spins arrange in a particular crystalline pattern with periodicity $p$ which maximizes their distance. The general formula for the unit cell defining this pattern is given~\cite{Hubbard78} in Eqs.~(\ref{Hubbard1}) and (\ref{Hubbard}). For example, for the fraction $q/p=2/5$ this gives $u_5 = \downarrow \uparrow \downarrow \downarrow \uparrow$.
Next we consider a transition between two crystalline phases $q/p \to q'/p'$ with $p,p' \le n+1$, and with unit cells $u_p$ and $u_{p'}$.
% Having defined the repeating strings $u_p$ and $u_{p'}$ for the fractions $q/p$ and $q'/p'$, respectively,
Consider the set of states formed by arbitrary sequences of the two unit cells, e.g., $...u_{p} u_{p} u_{p'} u_{p} u_{p}...$. We will show that this set of states is the degenerate ground state manifold at the classical $q/p \to q'/p'$ transition. Furthermore, associating the two types of unit cells  $u_{p}$ and $u_{p'} $ with either particles or holes living on a fictitious lattice (e.g., the state $u_{p} u_{p} u_{p'} u_{p} u_{p}$ is denoted as $00100$), we calculate the particle hopping amplitude (e.g., the amplitude for $00100 \to 00010$) at finite $J_{\perp}$. The result, leading to Eq.~(\ref{effgeneralized}), is that the hopping of the fictitious particles is obtained by means of a single operation of $J_\perp$ on the original spins.

\emph{Definition of the unit cell $u_p$ in a crystalline configuration:} Consider the crystalline state with fraction  $q/p$ of up spins. Its unit cell of length $p$, denoted $u_{p}$, is found using the following prescription:~\cite{Hubbard78} (i) define the integers $k,n_0,n_1,...,n_k$ by
\bea
\label{Hubbard1}
p/q &=& n_0+r_0,  \nonumber \\
|1/r_0|&=& n_1 + r_1,~~~|1/r_1|= n_2 + r_2,...\nonumber \\
|1/r_{k-2}|&=& n_{k-1} + r_{k-1},~~~|1/r_{k-1}| = n_k,
\eea
where for all $s$, $-\frac{1}{2} < r_s \le \frac{1}{2}$. (ii) Define the sequences $X_1,X_2,...,X_{k+1}$ and $Y_1,Y_2,...,Y_k$ by
\bea
\label{Hubbard}
X_1 &=& n_0, ~~~~~~~~~
Y_1 = n_0+\alpha_0,\nonumber \\
X_{i+1} &=& [X_i]^{n_i-1} Y_i ,~~
Y_{i+1} = [X_i]^{n_i+\alpha_i-1} Y_i,
\eea
where $\alpha_i = r_i / |r_i| = \pm 1$. (iii) Then, the required unit cell is $u_{p} = X_{k+1}$.

In a sense  $k$ measures the degree of ``fractallity" of the crystalline state: for $k=0$ unit cells have the form $X_1 = n_0$, consisting of $n_0-1$ down spins followed by one up spin, e.g., $X_1 = 3 \equiv \downarrow \downarrow \uparrow$.  For $k=1$, unit cells  have the form $X_{2} = [X_1]^{n_1-1} Y_1$, consisting of $n_1-1$ repetitions of the string $X_1$ followed by the string $Y_1 = X_1 \pm 1$, e.g., $ [3]^3[4]  = [3][3][3][4] $  $\equiv \downarrow \downarrow \uparrow \downarrow \downarrow \uparrow \downarrow \downarrow \uparrow \downarrow \downarrow \downarrow \uparrow$. Similarly, for a given $k$ the unit cells  $X_{k+1}$ consist of a number of repetitions of the string $X_{k}$ followed by the string $Y_k$.  To illustrate this construction we display in table \ref{table} all fractions and corresponding unit cells for $n=13$.
\begin{table}
\begin{center}
\begin{tabular}{|@{$\quad$}r@{$\quad$}r@{$\quad$}r@{$\quad$}r@{$\quad$} r@{$\quad$} |}
\hline
$\frac{q}{p}$ & period & $X_{k+1}=u_{p}$ & unit cell & $k$ \\
\hline
$\frac{1}{14}$ & 14 & $[14]$ & $ \downarrow \downarrow \downarrow \downarrow \downarrow \downarrow \downarrow \downarrow \downarrow \downarrow \downarrow \downarrow \downarrow \uparrow$ & 0 \\
$\frac{1}{13}$ & 13 & $[13]$ & $  \downarrow \downarrow \downarrow \downarrow \downarrow \downarrow \downarrow \downarrow \downarrow \downarrow \downarrow \downarrow \uparrow$ & 0 \\
$\frac{1}{12}$ & 12 & $[12]$ & $ \downarrow  \downarrow \downarrow  \downarrow \downarrow \downarrow \downarrow \downarrow \downarrow \downarrow \downarrow \uparrow$ & 0 \\
$\frac{1}{11}$ & 11 & $[11]$ & $ \downarrow \downarrow  \downarrow \downarrow \downarrow \downarrow \downarrow \downarrow \downarrow \downarrow \uparrow$ & 0 \\
$\frac{1}{10}$ & 10 & $[10]$ & $ \downarrow  \downarrow \downarrow \downarrow \downarrow \downarrow \downarrow \downarrow \downarrow \uparrow$ & 0 \\
$\frac{1}{9}$ & 9 & $[9]$ & $  \downarrow \downarrow \downarrow \downarrow \downarrow \downarrow \downarrow \downarrow \uparrow$  & 0\\
$\frac{1}{8}$ & 8 & $[8]$ & $  \downarrow \downarrow \downarrow \downarrow \downarrow \downarrow \downarrow \uparrow$ & 0 \\
$\frac{1}{7}$ & 7 & $[7]$ & $  \downarrow \downarrow \downarrow \downarrow \downarrow \downarrow \uparrow$ & 0 \\
$\frac{2}{13}$ & 13 & $[6][7]$ & $   \downarrow \downarrow \downarrow \downarrow \downarrow \uparrow \downarrow \downarrow \downarrow \downarrow \downarrow \downarrow \uparrow $ & 1 \\
$\frac{1}{6}$ & 6 & $[6]$ & $  \downarrow \downarrow \downarrow \downarrow \downarrow \uparrow $ & 0 \\
$\frac{2}{11}$ & 11 & $[5][6]$ & $    \downarrow \downarrow \downarrow \downarrow \uparrow \downarrow \downarrow \downarrow \downarrow \downarrow \uparrow $ & 1 \\
$\frac{1}{5}$ & 5 & $[5]$ & $   \downarrow \downarrow \downarrow \downarrow \uparrow $ & 0 \\
$\frac{3}{14}$ & 14 & $[4][5]^2$ & $ \downarrow \downarrow \downarrow \uparrow \downarrow \downarrow \downarrow \downarrow \uparrow \downarrow \downarrow \downarrow \downarrow \uparrow $ & 1 \\
$\frac{2}{9}$ & 9 & $[4][5]$ & $  \downarrow \downarrow \downarrow \uparrow \downarrow \downarrow \downarrow \downarrow \uparrow $ & 1 \\
$\frac{3}{13}$ & 13 & $[4]^2[5]$ & $   \downarrow \downarrow \downarrow \uparrow \downarrow \downarrow \downarrow \uparrow \downarrow \downarrow \downarrow \downarrow \uparrow $ & 1 \\
$\frac{1}{4}$ & 4 & $[4]$ & $    \downarrow \downarrow \downarrow \uparrow $ & 0 \\
$\frac{3}{11}$ & 11 & $[3][4]^2$ & $ \downarrow \downarrow  \uparrow \downarrow \downarrow \downarrow \uparrow \downarrow \downarrow \downarrow \uparrow$ & 1 \\
$\frac{2}{7}$ & 7 & $[3][4]$ & $  \downarrow \downarrow \uparrow \downarrow \downarrow \downarrow \uparrow$ & 1 \\
$\frac{3}{10}$ & 10 & $[3]^2[4]$ & $ \downarrow  \downarrow \uparrow \downarrow \downarrow \uparrow \downarrow \downarrow \downarrow \uparrow$ & 1 \\
$\frac{4}{13}$ & 13 & $[3]^3[4]$ & $  \downarrow \downarrow \uparrow \downarrow \downarrow \uparrow \downarrow \downarrow \uparrow \downarrow \downarrow \downarrow \uparrow$ & 1 \\
$\frac{1}{3}$ & 3 & $[3]$ & $    \downarrow \downarrow \uparrow $ & 0  \\
$\frac{5}{14}$ & 14 & $[3]^4 [2]$ & $ \downarrow \downarrow \uparrow \downarrow \downarrow \uparrow \downarrow \downarrow \uparrow \downarrow \downarrow \uparrow \downarrow \uparrow$  & 1 \\
$\frac{4}{11}$ & 11 & $[3]^3 [2]$ & $ \downarrow \downarrow  \uparrow \downarrow \downarrow \uparrow \downarrow \downarrow \uparrow \downarrow \uparrow$ & 1 \\
$\frac{3}{8}$ & 8 & $[3]^2 [2]$ & $  \downarrow \downarrow \uparrow \downarrow \downarrow \uparrow \downarrow \uparrow$ & 1 \\
$\frac{5}{13}$ & 13 & $[3]^2 [2] [3][2]$ & $  \downarrow \downarrow \uparrow \downarrow \downarrow \uparrow \downarrow \uparrow \downarrow \downarrow \uparrow \downarrow \uparrow$ & 2 \\
$\frac{2}{5}$ & 5 & $[2][3]$ & $   \downarrow \uparrow \downarrow \downarrow \uparrow $ & 1 \\
$\frac{5}{12}$ & 12 & $[2][3] [2]^2 [3]$ & $ \downarrow  \uparrow \downarrow  \downarrow \uparrow \downarrow \uparrow \downarrow \uparrow \downarrow \downarrow \uparrow$ & 2 \\
$\frac{3}{7}$ & 7 & $[2]^2 [3]$ & $  \downarrow \uparrow \downarrow \uparrow \downarrow \downarrow \uparrow$ & 1 \\
$\frac{4}{9}$ & 9 & $[2]^3[3]$ & $  \downarrow \uparrow \downarrow \uparrow \downarrow \uparrow \downarrow \downarrow \uparrow$  & 1\\
$\frac{5}{11}$ & 11 & $[2]^4 [3]$ & $ \downarrow \uparrow  \downarrow \uparrow \downarrow \uparrow \downarrow \uparrow \downarrow \downarrow \uparrow$ & 1\\
$\frac{6}{13}$ & 13 & $[2]^5 [3]$ & $  \downarrow \uparrow \downarrow \uparrow \downarrow \uparrow \downarrow \uparrow \downarrow \uparrow \downarrow \downarrow \uparrow$ & 1 \\
$\frac{1}{2}$ & 2 & $[2]$ & $    \downarrow \uparrow $ & 0 \\
\hline
\end{tabular}
\end{center}
\caption{Crystalline states stabilized for a finite range interaction Eq.~(\ref{frInt}) with $n=13$.
} \label{table}
\end{table}

\emph{Degenerate unit cells $u_p$ and $u_{p'}$ at classical transitions:}
It is elementary to show that at transitions [between closest fractions $0 \le q/p < q'/p' \le 1$ with $p,p' \le n+1$] $k$ either (i) remains constant, $k=k'$, or (ii)  $k - k' = \pm 1$; see table \ref{table}.
 For transitions of type (i) we can write
    \be
    \label{kequal}
    \{ u_{p}, u_{p'} \}= \{X_{k''}, Y_{k''} \},
     \ee
     where $\{X_{k''}, Y_{k''} \}$ are given from Eq.~(\ref{Hubbard}) evaluated for a third fraction, $q''/p''=(q+q')/(p+p')$ with degree of ``fractallity" $k''$ (that has the unit cell $X_{k''} Y_{k''}$ which is not stabilized for the considered interaction range); For transitions of type (ii) we may write   $\{ u_{p}, u_{p'} \}= \{X_{k+1}, X_k \}$, where $ \{X_{k+1}, X_k \}$ are given from Eq.~(\ref{Hubbard})  evaluated for the (smaller) fraction $q/p$. [in case (ii), it is sufficient to consider the case that $k'= k-1$].

\emph{Degeneracy of the sub-Hilbert space:}  We associate  $u_{p}$ ($u_{p'}$) with particles (empty sites) on a fictitious lattice and set the magnetic field $\Delta$ to the transition between the corresponding commensurate crystalline phases. We need to show that at  $J_\perp=\Omega=0$, states with arbitrary sequences of $u_{p}$ and $u_{p'}$, corresponding to arbitrary particle states in the fictitious lattice, are degenerate.
Since the condition of convexity is assumed, upon increasing $\Delta$ all sates with rational density of up spins become a ground state (possibly degenerate) at some field. If the interaction is of infinite range than every fraction acquires a plateau.~\cite{Bak82} For finite range interaction,  however, the two crystalline states $... u_{p} u_{p} u_{p} u_{p}...$ (maximal particle  occupation) and $... u_{p'} u_{p'} u_{p'} u_{p'}...$ (zero particle occupation) become degenerate at their transition field also with an infinite set of crystalline phases, e.g.,
$... u_{p} u_{p'} u_{p} u_{p'}...$. The unit cells of such periodic arrangement correspond via Eqs.~(\ref{Hubbard1}) and (\ref{Hubbard}) to  fractions in the open interval $( q/p ,q'/p') $ which might have arbitrary large $k$ values; these fractions necessarily have a denominator $ > n+1$ and  hence are not stabilized over a finite range of longitudinal magnetic field $\Delta$. Hence, for any  density of particles there exist  degenerate   particle states. For example, if $k = k'$, then we can construct states with arbitrary low particle density of the form $u_p (u_{p'})^n$ with any integer $n$, or states with arbitrarily large particle density $\to 1$ of the form $(u_p)^n u_{p'}$. However, those states are very special as they correspond to some periodic crystalline arrangement of the spins.

To show that \emph{all} particle configurations are degenerate, it remains to be shown that one can start with one of those periodic states and move particles around without any energy change. The basic such  process that we need to consider is the single particle hopping, $... u_{p'} u_{p} ... \to ... u_{p} u_{p'} ...$.
We will consider transitions of type (i)   between two states with the same value of $k$ where $u_p=X_{k''}$ and $u_{p'}=Y_{k''}$ with $k''=k+1=k'+1$; see Eq.~(\ref{kequal}). Case (ii) can be addressed along the same lines. We start with the initial configuration
\bea
\label{initial}
|i \rangle =...Y_{k+1} X_{k+1}...=...0 1 ...,
\eea
which contains a $Y_{k+1}$ string followed by an $X_{k+1}$ string, surrounded by a particular configuration of either one of those strings, hidden in the $(...)$ which is fixed to define the state, but can be arbitrary, and we would like to arrive at the final state
\bea
\label{final}
|f \rangle =...X_{k+1} Y_{k+1}...=... 1 0 ...,
\eea
which coincides with the initial state except for the exchange $Y_{k+1} X_{k+1} \to X_{k+1} Y_{k+1}$. Our purpose below, is to show that  this exchange process consists of a single operation of $J_\perp$, and that it does not cost energy, independent of the unspecified (...) configurations.

We use Eq.~(\ref{Hubbard}) to express $X_{k+1}$ and $Y_{k+1}$ in terms of $X_{k}$ or $Y_{k}$. We obtain
\bea
\label{intermediate}
| i \rangle = ... [X_{k}]^{m_{k}} Y_{k} X_{k} [X_{k}]^{m_{k}} Y_{k}..., \nonumber \\
| f \rangle = ... [X_{k}]^{m_{k}} X_{k} Y_{k} [X_{k}]^{m_{k}} Y_{k}...,
\eea
for $\alpha_{k} = -1$, or
\bea
| i \rangle = ... [X_{k}]^{m_{k}} X_{k} Y_{k} [X_{k}]^{m_{k}} Y_{k}..., \nonumber \\
| f \rangle = ... [X_{k}]^{m_{k}} Y_{k} X_{k} [X_{k}]^{m_{k}} Y_{k}...,
\eea
for $\alpha_{k} = 1$, where $m_i=n_i-1+\frac{\alpha_i-1}{2} = \min \{n_i-1 , n_i -1 +\alpha_i \}$. Using Eq.~(\ref{Hubbard}) we continue successively decomposing $X_{k}$ and $Y_{k}$  into $X_{k-1}$ and $Y_{k-1}$ until we reach the $X_1$ and $Y_1$ unit cells. The result is
\bea
\label{if3}
| i \rangle =... B Y_1 X_1 A...,     \nonumber \\
| f \rangle =... B X_1 Y_1 A...,
\eea
for $\prod_{i=1}^{k}  (- \alpha_i) = 1$, or
\bea
\label{if4}
| i \rangle = ...B X_1 Y_1 A...,     \nonumber \\
| f \rangle = ...B Y_1 X_1 A...,
\eea
for $\prod_{i=1}^{k}  (- \alpha_i) = 1$, where
\bea
A &=&\prod_{i=1}^{k} [X_i]^{m_i}[Y_i], \nonumber \\
B &=& \prod_{i=0}^{k-1}[X_{k-i}]^{m_{k-i}}.
\eea
Furthermore, $Y_1 X_1 = [\downarrow]^{m_{0}}(\uparrow \downarrow)[\downarrow]^{m_{0}}\uparrow$ if $\alpha_0 = -1$, or $Y_1 X_1 = [\downarrow]^{m_{0}} ( \downarrow \uparrow) [\downarrow]^{m_{0}}\uparrow$ if $\alpha_0 = 1$, hence we obtain
\bea
\label{if1}
| i \rangle =...B   [\downarrow]^{m_{0}}(\uparrow \downarrow)[\downarrow]^{m_{0}}\uparrow  A..., \nonumber \\
| f \rangle =...B   [\downarrow]^{m_{0}} ( \downarrow \uparrow) [\downarrow]^{m_{0}}\uparrow  A...,
\eea
for $\prod_{i=0}^{k}  (- \alpha_i) = 1$, or
\bea
\label{if2}
| i \rangle =...B   [\downarrow]^{m_{0}} ( \downarrow \uparrow) [\downarrow]^{m_{0}}\uparrow  A..., \nonumber \\
| f \rangle =...B     [\downarrow]^{m_{0}}(\uparrow \downarrow)[\downarrow]^{m_{0}}\uparrow  A ...,
\eea
for $\prod_{i=0}^{k}  (- \alpha_i) = 1$.
From Eqs.~(\ref{if1}) and (\ref{if2}) we see that
\bea
\langle i | \mathcal{H}  | f \rangle =\langle f | \mathcal{H}  | i \rangle= -\frac{J_\perp}{2},
\eea
where $\mathcal{H}$ is given in Eq.~(\ref{Model}), namely, the hopping amplitude is just that of flipping the two spins in parentheses, achieved by a single operation of $J_\perp$. We proceed with the proof that the interaction energy of the initial and final states are equal.

The unspecified $(...)$ to the right in Eqs.~(\ref{initial}) and (\ref{final}), either starts with $ X_{k+1} = [X_{k}]^{n_{k}-1} Y_{k}$ or $Y_{k+1} = [X_{k}]^{n_{k}-1+\alpha_{k}} Y_{k}$, therefore  it begins with $[X_{k}]^{m_{k}}$. Similarly, the undetermined $(...)$ to the left must start with $[X_{k}]^{m_{k}}Y_{k}$. Thus, we can display more spins for the same state given in Eq.~(\ref{intermediate}),
\bea
| i \rangle &=& ...[X_{k}]^{m_{k}}Y_{k} [X_{k}]^{m_{k}} ( X_{k} Y_{k})
\nonumber \\
&& [X_{k}]^{m_{k}} Y_{k}   [X_{k}]^{m_{k}}...
\eea
(for $\alpha_{k}=-1$). We may therefore successively introduce in the right side $(...)$ in Eqs.~(\ref{if1}) and (\ref{if2}) the combinaiton $[X_{k-1}]^{m_{k-1}}$, then $[X_{k-2}]^{m_{k-2}}$, till $[X_{1}]^{m_{1}}$. Similarly, we successively add in the (...) to the left the strings $[X_{k-1}]^{m_{k-1}}Y_{k-1}$, then $[X_{k-2}]^{m_{k-2}}Y_{k-2}$, till $[X_{1}]^{m_{1}}Y_{1}$. The result, for $\prod_{i=1}^{k}  (- \alpha_i) = 1$, is
\bea
\label{if3a}
| i \rangle =... AB  [\downarrow]^{m_{0}}(\uparrow \downarrow)[\downarrow]^{m_{0}}\uparrow AB...,     \nonumber \\
| f \rangle =... AB [\downarrow]^{m_{0}}(\downarrow \uparrow)[\downarrow]^{m_{0}}\uparrow AB....
\eea
The unspecified $(...)$ to the right in Eqs.~(\ref{if3a}), either starts with $ X_1 = [\downarrow]^{n_0-1} \uparrow$ or $Y_1 = [\downarrow]^{n_0-1+\alpha_0} \downarrow$, hence it begins with $[\downarrow]^{m_{0}}  = \underbrace{\downarrow \downarrow ... \downarrow}_{m_0}$. Similarly, the unspecified $(...)$ to the left must end with $[\downarrow]^{m_{0}} \uparrow$.
Hence we obtain
\bea
\label{if1a}
| i \rangle =...C (\uparrow \downarrow)  C ..., \nonumber \\
| f \rangle =...C  (\downarrow \uparrow)   C ...,
\eea
for $\prod_{i=0}^{k}  (- \alpha_i) = 1$, or
\bea
\label{if2a}
| i \rangle =...C (\downarrow \uparrow)  C ..., \nonumber \\
| f \rangle =...C  (\uparrow \downarrow)    C ...,
\eea
for $\prod_{i=0}^{k}  (- \alpha_i) = 1$. Here,
\bea
\label{C}
C = [\downarrow]^{m_{0}} \uparrow AB [\downarrow]^{m_{0}} .
\eea
What we have done in Eqs.~(\ref{if1a}) and (\ref{if2a}) is to maximally determine the spins to the left and right of the exchanged up and down spins shown in parentheses, which  physically follows from the crystalline underlying interaction. The interaction of the up spin in parentheses with all other up spins within the string displayed in Eqs.~(\ref{if1a}) and (\ref{if2a}) is equal for the final and initial states. This follows because  the string $C$ is related to itself by left-right inversion. This property, $C = \bar{C}$, where $\bar{X}$ is the left-right inversion of a string $X$, follows from the induction proven set of identities
\be
[\uparrow] \prod_{j=1}^{i-1} Y_j X_i = \bar{X}_i \prod_{j=1}^{i-1} \bar{Y}_{i-j} [\uparrow],
\ee
\be
 [\uparrow] \prod_{j=1}^{i} Y_j = \prod_{j=1}^{i} \bar{Y}_{i+1-j} [\uparrow],
\ee
for any $i \ge 1$.
 Furthermore, due to the finite interaction range, $J^{(n)}_{|r|}=0$ for $|r|>n$, the up spin in parentheses does not interact with any spin out of the displayed sequence in Eqs.~(\ref{if1}) and (\ref{if2}), as we now demonstrate. Denoting the total number of spins (up or down) in a string  $X$ by $L_{X}$, we have $|r|\ge L_C+1 $. Furthermore, $L_C =  L_{[X_{k+1}][Y_{k+1}]}-2$, which follows by comparing Eqs.~(\ref{if1}), (\ref{if2}), and (\ref{C}). Since a crystalline state with unit cell $[X_{k+1}][Y_{k+1}]$, corresponding to fraction $(q+q')/(p+p')$,  is not stabilized, we have  $L_{[X_{k+1}][Y_{k+1}]} \ge n+2$. This implies $|r| \ge n+1$ hence $J^{(n)}_{|r|}=0$, completing our proof that the interaction energy in the initial and final states are equal.

\end{document}